\documentclass[aps,prb,reprint,showpacs,superscriptaddress]{revtex4-2}

\usepackage{graphicx}
\usepackage{dcolumn}
\usepackage{bm}
\usepackage{color}  
\expandafter\let\csname equation*\endcsname\relax
\expandafter\let\csname endequation*\endcsname\relax
\usepackage{amsmath}
\usepackage{graphicx}
\usepackage{soul}
\usepackage{float}
\emergencystretch=\maxdimen
\usepackage[export]{adjustbox}
\DeclareMathOperator{\csch}{csch}

\begin{document}

\title{Brownian Fluctuations of a non-confining potential}

\author{Pedro B. Melo}
 \affiliation{Departamento de F\'isica, Pontif\'icia Universidade Cat\'olica, CEP 22452-970, Rio de Janeiro, Brazil}
\author{Pedro V. Paraguass\'u}%
\affiliation{Departamento de F\'isica, Pontif\'icia Universidade Cat\'olica, CEP 22452-970, Rio de Janeiro, Brazil}%
\author{Eduardo S. Nascimento}
\affiliation{Departamento de F\'isica, Universidade Federal do Piau\'i, CEP 64049-550, Teresina, Piau\'i, Brazil}
\author{Welles A. M. Morgado}
\email{welles@puc-rio.br}
\affiliation{Departamento de F\'isica, Pontif\'icia Universidade Cat\'olica, CEP 22452-970, Rio de Janeiro, Brazil}%

\date{\today}

\begin{abstract}
Brownian fluctuations arise for any quantity that depends on the stochastic variables of a Brownian particle. In this study, we explore the Brownian fluctuations of a bidimensional quadratic potential that exhibits two regimes: a confining regime and a non-confining regime. We divide the total potential into two contributions and analyze the mean, variance, skewness, excess kurtosis, and their distributions for each contribution as well as for the total potential. Our analysis offers an understanding into the statistical behavior of each quantity.
\end{abstract}

\maketitle

\onecolumngrid
\section{Introduction}

Out of equilibrium systems display novel phenomena due to the relevance of fluctuations. One of the simplest systems of this kind is the Brownian particle \cite{Sekimoto2010}. It is the minimal continuous system exhibiting diffusion and the influences of thermal fluctuations. The behavior of this kind of system is described not only by averages, but also by probability distributions \cite{tome2015stochastic}.

In the field of Stochastic Thermodynamics \cite{oliveira_2020,peliti2021stochastic}, distinct  thermodynamic quantities such as heat, work, and entropy are linked to the behavior of Brownian particles \cite{Sekimoto2010}. These quantities fluctuate stochastically, as does the position of the particle itself. These Brownian fluctuations in these thermodynamic quantities have been extensively studied both theoretically \cite{salazar2019stochastic, salazar2020work, colmenares2021optimal, colmenares2022work, chatterjee2011single,chatterjee2010exact, saha2014work, pires2023optimal,paraguassu2023heat2,Paraguassú_2021,paraguassu2022effects,paraguassu2023heat} and experimentally in a variety of systems \cite{darabi2023stochastic,gomez2021work,imparato2008probability,joubaud2007fluctuation,ciliberto2017experiments}. They exhibit non-trivial statistical behaviors that go beyond the typical Gaussian distribution \cite{paraguassu2023heat}.

Taking this approach into consideration, our current interest lies in investigating the Brownian fluctuations in the change of potential energy for a potential that has two distinct regimes: one where it confines motion, and another where it doesn't, depending on specific parameters of the potential. It's worth mentioning that one of the contributions of the heat is the change in potential energy, which has been previously explored in prior research \cite{chatterjee2010exact,chatterjee2011single,Paraguassú_2021}. However, even in the overdamped limit, it's essential to recognize that heat is always associated with kinetic energy, as demonstrated in \cite{paraguassu2022effects}.

The introduction of non-confining potentials leads to non-equilibrium states where the particle does not achieve a Boltzmann distribution after long times. This phenomena is associated with a divergence of quantities such as the variance of the probability distribution for physical quantities of interest such as the position and energy. Here, we are interested in a two-dimensional overdamped Brownian particle, subject to a quadratic potential, $V(x,y)= {k}(x^2+y^2)/2-uxy$ (see \ref{potential}), which induces correlations between the particle's positions in both the $x$ and $y$ directions, and is non-confining for $u=k$. It is worth to mention that a particular stochastic system with quadratic potential that has been received a lot of interest is the Brownian Gyrator \cite{Gyrator1}, which exhibits a non-equilibrium steady state (NESS), with complex gyrating properties, due the influence of distinct thermal baths and anisotropic terms of the potential \cite{ArgunVolpe2017,MancoisWilkowski2018,dos2021stationary}. However, in the case $k\rightarrow u$, we shall confine our study to a single temperature, which does not exhibits a NESS, but a non-equilibrium trivial state that fails to achieve a Boltzmann distribution of a particle that has a ballistic expansion \cite{vk}, where the particle behaves as a free particle.

Moreover, for systems under non-confining potentials in general, many aspects were already analyzed, such as the free energy of a particle within a square potential well \cite{Farago2021}, where its free energy behaves similarly to the free energy of a free particle diffusion. The work fluctuation theorem for a Brownian particle under non-confining potential was also calculated \cite{Kantz2021}. Systems with non-confining potentials such as the logarithm potential have been extensively analyzed in the literature, for instance to calculate the probability density function of an ensemble of Brownian particles under this kind of potential \cite{FokkerPlanck_Logarithmic} and the heat and work functional for a Brownian particle in the same regime \cite{Paraguassú_2021, ryabov2013work}. Besides these calculations, there is also an effort in the direction of regularization of the Boltzmann-Gibbs statistics of a system under non-confining potentials \cite{Aghion2019_prl,AGHION2020109890,Defaveri2020}.

In this paper, we analyze the Brownian fluctuations of the  potential energy increment by means of path integrals \cite{WioPath1}. We investigate each term that composes the potential energy and the total potential energy separately. We calculate the characteristic function, its associated central moments and  the probability distribution for the harmonic, non-confining component and the total potential energy of the system. Comparing the results with Langevin equation simulations, we find a good agreement. 

This paper is organized as follows: In Sec. \ref{model}, we introduce the model of a two-dimensional Brownian particle in a non-confining potential and the method of solving the Langevin equation, by stochastic path integrals. In Sec.~\ref{harmonic}, we present the fluctuations for the harmonic potential. In Sec.~\ref{nonconfining}, we present the fluctuations for the non-confining potential. In Sec.~\ref{fullpotential}, we present the fluctuations for the full potential and make the connection between our results and the fluctuations of the heat. Finally, in Sec.~\ref{Discussion}, we present our discussions and our conclusions.

\section{\label{model}Model and Dynamics}

\begin{figure}
    \centering
    \includegraphics[width=15cm]{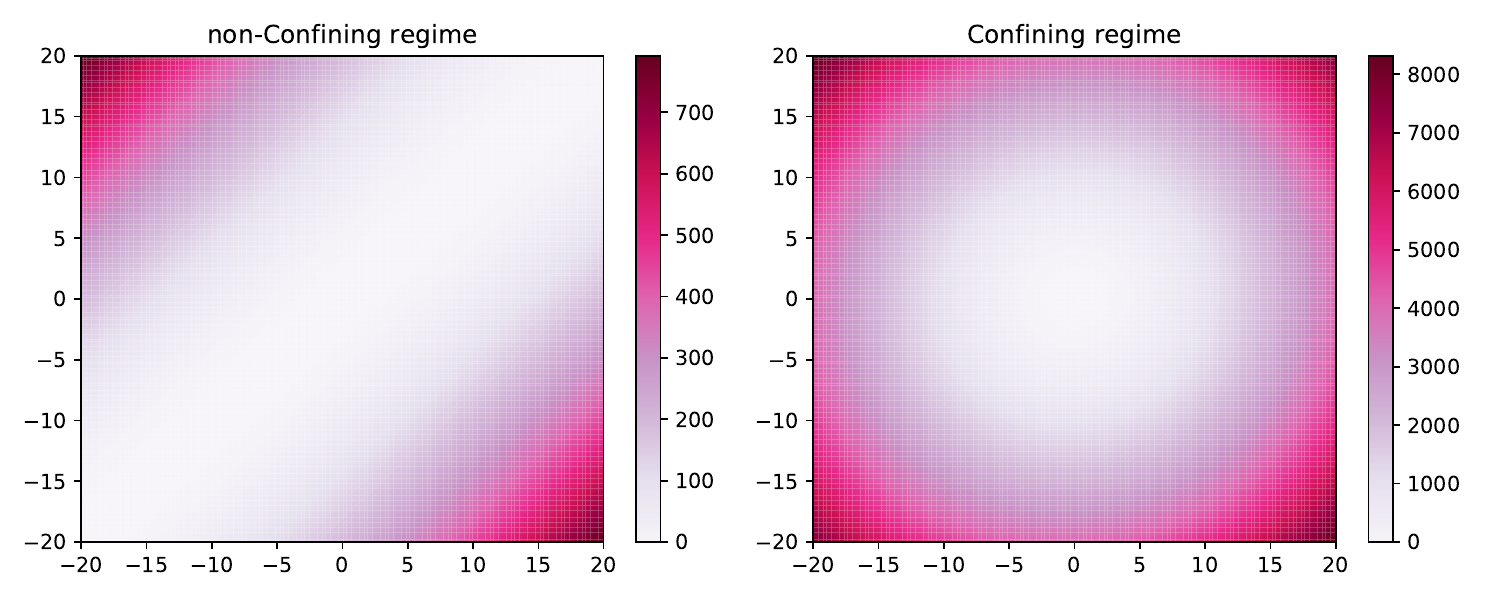}
    \caption{Two distinct regimes of the full potential, $V(x,y)$, in the left the non-confining, where $k=u=1$, and in the right the confining, for $k=10,u=1$. Note that, despite the coupling, the confining is qualitatively almost of the same shape of the harmonic potential. } 
    \label{potential}
\end{figure}

The adopted model is a two-dimensional overdamped  Brownian particle, given by the set of Langevin equations
\begin{eqnarray}
    \gamma\dot{x}(t) = -kx(t) + uy(t) +\eta_{x}(t),\\
    \gamma\dot{y}(t) = -ky(t) + ux(t) +\eta_{y}(t), \label{langevin}
\end{eqnarray}
where $x(t)$ and $y(t)$ are the positions of the particle, $\gamma$ is the damping constant that refers to friction that a particle suffers, $k$ the spring constant due the harmonic potential, $u$ a constant that couples both $x$ and $y$, the strength of the torque, and $\eta_{i}(t)$ ($i = x, y$) refers to the noise term which we consider to be a white noise, with $\langle \eta_{i}(t)\rangle = 0$ and $\langle \eta_{i}(t) \eta_{j}(t')\rangle = 2\gamma T\delta_{ij}\delta(t - t')$.\\ 

Here we are interested in the fluctuations of distinct potential terms. In our model we have a main contribution coming from a harmonic potential 
\begin{equation}
    U= \frac{k}{2} \left(x^2+y^2\right),
\end{equation}
which is a confining potential, and could represent a trapped particle in two dimensions. Also, we consider the coupling energy

\begin{equation}
    V_{nc} = -uxy,
\end{equation}

which may promote a confining or non-confining behavior, depending on the relation between $k$ and $u$. As a result, the total potential reads
\begin{equation}
    V = U + V_{nc}.
\end{equation} 

The force defined in Eq.~\ref{langevin} is the derivative of $V$,  $\Vec{F}= -\nabla V$. This contribution affect the shape of the potential depending on the value of $k$ and $u$. For values of $k$ greater than $u$, we observe a well-type potential. However, it's important to emphasize that, due to the interaction term $u$, we introduce a coupling between the position variables that affect the dynamics. Furthermore, for $k=u$, we have non-confining behavior, where the particle's equilibrium is no longer achievable, and the potential become $V= \frac{k}{2}(x-y)^2$. The confining and non-confining versions of this potential are shown in  Fig. \ref{potential}. 

Besides that, the two positions can be interpreted as two harmonic oscillators interacting via the non-confining contribution, which resembles to the coherent scattering interaction well studied in the context of quantum optomechanics~\cite{brandao2021coherent, ranfagni2021vectorial}.

The Brownian Gyrator presents a quadratic potential energy but, in order to exhibit gyrating properties \cite{Gyrator1,dos2021stationary}, two distinct temperatures are required. In this analysis, we shall only consider a single bath. Nonetheless, we will observe that we can still derive intriguing findings regarding the non-confining potential energy of this system.

By means of path integral calculations, one can calculate the conditional probability $P(x_\tau,y_\tau|x_{0},y_{0})$, given by~\cite{WioPath1}
\begin{equation}
P(x_\tau,y_\tau|x_0,y_0) = \int \mathcal{D}x\int \mathcal{D}y 
    \exp\left(-\frac{1}{4\gamma T}S[x,y]\right),
\end{equation}
where $S[x,y]$ is the stochastic action, given by
$$ S[x,y] = \int_0^\tau \left(\left(\gamma \dot x + k x + u y\right)^2 +\left(\gamma \dot y + k y + u x\right)^2\right)dt. $$
The calculation  of the conditional probability is detailed shown in \ref{pathcalculations}. The solution is a correlated Gaussian distribution for the variables $x_\tau$, $y_\tau$, $x_{0}$, and $y_{0}$:
\begin{eqnarray}
 P(x_\tau, y_\tau | x_0, y_0) = \frac{1}{4 \pi T}\sqrt{(k^2-u^2)  \left(\coth \left(\frac{\tau (k-u)}{\gamma}\right)+1\right) \left(\coth \left(\frac{\tau (k+u)}{\gamma}\right)+1\right)} \nonumber\\ 
\times \exp \Bigg\{-\frac{1}{4 \gamma T}(x_{0}^2+y_{0}^2) \left( -\gamma k - \frac{1}{4} \gamma \left((k-u) \cosh \left(\frac{\tau (k-u)}{\gamma}\right) - k \cosh \left(\frac{\tau (3k+u)}{\gamma}\right) \right.\right.\nonumber\\ \left.\left. + u \cosh \left(\frac{\tau (k+3u)}{\gamma}\right)\right)\csch\left(\frac{\tau (k-u)}{\gamma}\right)\csch^2\left(\frac{\tau (k+u)}{\gamma}\right) \right) + \frac{1}{2} \gamma \csch\left(\frac{\tau (k-u)}{\gamma}\right)\nonumber\\ \times\left( -\left((k+u) (x_{0}-y_{0}) (x_{\tau}-y_{\tau}) \left(\cosh \left(\frac{2k\tau}{\gamma}\right) -\cosh\left(\frac{2\tau u}{\gamma}\right)\right) \csch^2\left(\frac{\tau (k+u)}{\gamma}\right)\right) \right) \nonumber\\
- 2 (k-u) (x_{0}+y_{0}) (x_{\tau}+y_{\tau}) + x_{0}y_{0} \left(2\gamma u - \frac{1}{2} \gamma \left((k-u) \cosh\left(\frac{\tau (k-u)}{\gamma}\right) + u \cosh\left(\frac{\tau (3k+u)}{\gamma}\right) \right.\right. \nonumber\\
\left. \left. - k \cosh\left(\frac{\tau (k+3u)}{\gamma}\right)\right) \csch\left(\frac{\tau (k-u)}{\gamma}\right)\csch^2\left(\frac{\tau (k+u)}{\gamma}\right) \right) + \frac{1}{2} \left(\gamma (k+u) (x_{\tau}-y_{\tau})^2 \right.\nonumber\\
\left. \times\left(\coth \left(\frac{\tau (k+u)}{\gamma}\right)+ 1\right) + \gamma (k-u) (x_{\tau}+y_{\tau})^2 \left(\coth\left(\frac{\tau (k-u)}{\gamma}\right) + 1\right)\right)\Bigg\}.
\end{eqnarray}
Considering a particle initially centered in the origin, with $x_{0} = 0$ and $y_{0} = 0$, the conditional probability is given by
\begin{eqnarray}
    P(x_\tau,y_\tau|x_0=0,y_0=0) = \sqrt{(k^2-u^2)  \left(\coth \left(\frac{\tau  (k-u)}{\gamma }\right)+1\right) \left(\coth \left(\frac{\tau  (k+u)}{\gamma }\right)+1\right)}\nonumber \\ \times \frac{1}{4 \pi  T} \exp \left(-\frac{\gamma  (k+u) (x_\tau-y_\tau)^2 \left(\coth \left(\frac{\tau  (k+u)}{\gamma }\right)+1\right)}{8 \gamma  T}\right) \nonumber \\  \times \exp \left(-\frac{\gamma  (k-u) (x_\tau+y_\tau)^2 \left(\coth \left(\frac{\tau  (k-u)}{\gamma }\right)+1\right)}{8 \gamma  T}\right).\label{conditional}
\end{eqnarray}
This conditional distribution informs us about the correlation between initial and final positions. With it, we can construct the joint distribution, which will be important for calculating the distribution of potentials. The joint probability is given by
\begin{equation}
    P(x_\tau,y_\tau,x_{0},y_{0}) = P(x_\tau,y_\tau|x_0,y_0)P_0(x_0,y_0),
    \label{jointdist}
\end{equation}
where $P_{0}(x_{0},y_{0}) = \delta(x_{0})\delta(y_{0})$, where $\delta(x_{0})$ is a delta function.  Observe that the initial condition adopted corresponds to a true zero entropy initial state. 

It is important to emphasize that the conditional probability is valid for any value of $k$ and $u$, including when $k<u$. For $k=u$, in the non-confining regime, we have 
\begin{eqnarray}
 P(x_{\tau}, y_{\tau} | x_{0}, y_{0})_{k\rightarrow u} = \frac{1}{2 \sqrt{2} \pi T} \sqrt{\frac{\gamma u (\coth(\frac{2 \tau u}{\gamma}) + 1)}{\tau}} \exp\left(-\frac{1}{8T} \left(2u\,\left((x_{0} - y_{0})^2 + (x_{\tau} - y_{\tau})^{2}\right)\,\right.\right.\nonumber \\
\times\coth\left(\frac{2 \tau u}{\gamma}\right) - 4u(x_{0} - y_{0})(x_{\tau} - y_{\tau})\csch\left(\frac{2 \tau u}{\gamma}\right) - 2u(x_{0} + x_{\tau} - y_{0} - y_{\tau})(x_{0} - x_{\tau} - y_{0} + y_{\tau}) \nonumber \\
 \left.\left. + \frac{\gamma(x_{0} - x_{\tau} + y_{0} - y_{\tau})^2}{\tau}\right)\right)
\end{eqnarray}
For the particle initially centered on the origin with $x_0 = 0$ and $y_0 = 0$, we have
\begin{eqnarray}
    P(x_\tau,y_\tau|x_0,y_0)_{k\rightarrow u} = \exp\left(-\frac{2 \tau  u (x_\tau-y_\tau)^2 \left(\coth \left(\frac{2 \tau  u}{\gamma }\right)+1\right)+\gamma  (x_\tau+y_\tau)^2}{8 \tau  T}\right) \nonumber \\ \times \frac{1}{2 \sqrt{2} \pi  T}\sqrt{\frac{\gamma  u \left(\coth \left(\frac{2 \tau  u}{\gamma }\right)+1\right)}{\tau }},
\end{eqnarray}
which is entirely finite, in accordance with Aghion \textit{et al.}~\cite{Aghion2019_prl}. Furthermore, we see that a temporal dependency arise within the argument of the $\coth$ (in fact, coupled to $u$), which is a discontinuous and divergent function at the origin. 

Nevertheless, for $k>u$, we expect the particle to evolve to an equilibrium state for asymptotic time, which can be seen by taking $\tau \rightarrow \infty$ in the distribution of Eq.~\ref{conditional} 
\begin{equation}
    \lim_{\tau \rightarrow\infty} P(x_\tau,y_\tau|x_0,y_0) \rightarrow \frac{\sqrt{k^2-u^2} }{2 \pi  T}\exp\left({-\frac{k \left(x_\tau^2+y_\tau^2\right)-2 u x_\tau y_\tau}{2 T}}\right).
\end{equation}
The Boltzmann distribution is observed in the system under consideration. Additionally, due to the coupling $u$, it is noteworthy that a correlation exists between the $x$ and $y$ components. When $u=0$, the equilibrium distribution is obtained for a particle confined within a harmonic potential characterized by a constant $k$.

However, when $k=u$, the distribution becomes null, thereby rendering the attainment of Boltzmann equilibrium infeasible, 
as expected for an unbounded distribution. A consequence of the potential being in the non-confining regime. Despite of that, we can still investigate the fluctuations of the harmonic potential and the full potential in both regimes, as we will do in the following sections.

\section{Harmonic Energy Fluctuations\label{harmonic}} 

The fluctuations inherent to Brownian motion give rise to statistical fluctuations in the potential energy of the particle. In this context, we proceed to calculate the fluctuations in the harmonic potential energy contribution to the total potential.  The variation in potential energy over a trajectory $t\in[0,\tau]$, can be expressed as:
\begin{equation}
\Delta U(x,y) = \frac{k}{2} \left(x_{\tau}^2+y_{\tau}^2-x_0^2-y_0^2\right).
\end{equation}
It is for this quantity that we will investigate its  distribution, and consequent fluctuations. To  obtain the fluctuations, our first step is to calculate the characteristic function, given by 
\begin{eqnarray}
    Z_{\Delta U}(\lambda) = \langle e^{-i\lambda \Delta U(x,y)} \rangle = \int dx_{\tau}dy_{\tau}dx_{0}dy_{0}\; e^{-i\lambda \Delta U(x,y)}P(x_{\tau},y_{\tau},x_{0},y_{0}).
\end{eqnarray}
Since, $P(\dots)$ are the product of a Gaussian joint distribution and a Dirac delta (as showed in Eq.\ref{A9}), and $\Delta U$ has a quadratic dependency on $x$ and $y$, its solution will be of the form
\begin{equation}
   Z_{\Delta U}(\lambda) = \frac{\sqrt{\alpha_3}}{\sqrt{\alpha_1\lambda^2 + \alpha_2\lambda + \alpha_3}},
\end{equation}
where, the coefficients of the characteristic function will be given by 
\begin{eqnarray}
    \alpha_{1} = -4k^2T^2,\\
    \alpha_{2} = (k^2 - u^2)\left(1 + f_C(k - u)\right)\left(1 + f_C(k + u)\right),\\
    \alpha_{3} = -2ikT\left( (u-k) f_C(k - u) - (k+u)f_C(k + u) - 2k \right),
\end{eqnarray}
where we define the quantities $f_C(k \pm u)$ to simplify the notation:
\begin{equation}
    f_C(k \pm u)=\coth \left(\frac{(k\pm u)\tau}{\gamma }\right). 
\end{equation}
In the remaining of the paper, due to their mathematical similarity, all the investigated quantities shall have the same expression for the characteristic function, with only the $\alpha_i$ coefficients changing.We wrote a short review about this quantity in \ref{appB}, along with the calculation of the probability distribution from the characteristic function, in cases where an analytical solution is possible.

\subsection{Central Moments}
 Cumulants, and central moments, may provide information about the statistical behavior of the random variable of interest. They can be obtained by taking derivatives of the characteristic function \cite{paraguassu2023heat} (see \ref{appB}). The moments and central moments constructed are: the mean, variance, skewness, and excess kurtosis. We investigate its asymptotic values and the behavior in the confining and non-confining regime.

The first moment of the non-interaction energy term, $U$, is its average, given by
\begin{equation}
    \mu_{\Delta U} \equiv \langle \Delta U \rangle = kT\left(\frac{1}{(k+u)(f_C(k + u) + 1)}+\frac{1}{(k-u) \left(f_C(k - u) +1\right)}\right).\label{mean_U}
\end{equation}
The average is monotonically  increasing in time, since $f_C(\dots)$ will be monotonically decreasing, and is always positive. For values $k>u$ the average saturates for $\tau\rightarrow\infty$, given by
\begin{equation}
    \lim_{\tau\rightarrow\infty}\langle \Delta U \rangle = \frac{k^2 T}{k^2-u^2},
\end{equation}
which is divergent for $k=u$, that is, in the non-confining regime. Naturally, the asymptotic limit gives a positive number, as the potential is quadratic. Nevertheless, for transient times, we can obtain an expression for the mean in the non-confining regime, which is
\begin{equation}
        \lim_{k\rightarrow u}\langle \Delta U \rangle = \frac{T}{4} \left(\frac{4 \tau  u}{\gamma }-e^{-\frac{4 \tau  u}{\gamma }}+1\right).
\end{equation}
Notice that for $\tau\rightarrow\infty$ this expression clearly diverges because of the linear dependence on $\tau$. It happens that the particle moves far from the origin $x=y=0$, along the line $x=y$. 

\begin{figure}[h]
    \centering
    \includegraphics[width=1.0\textwidth]{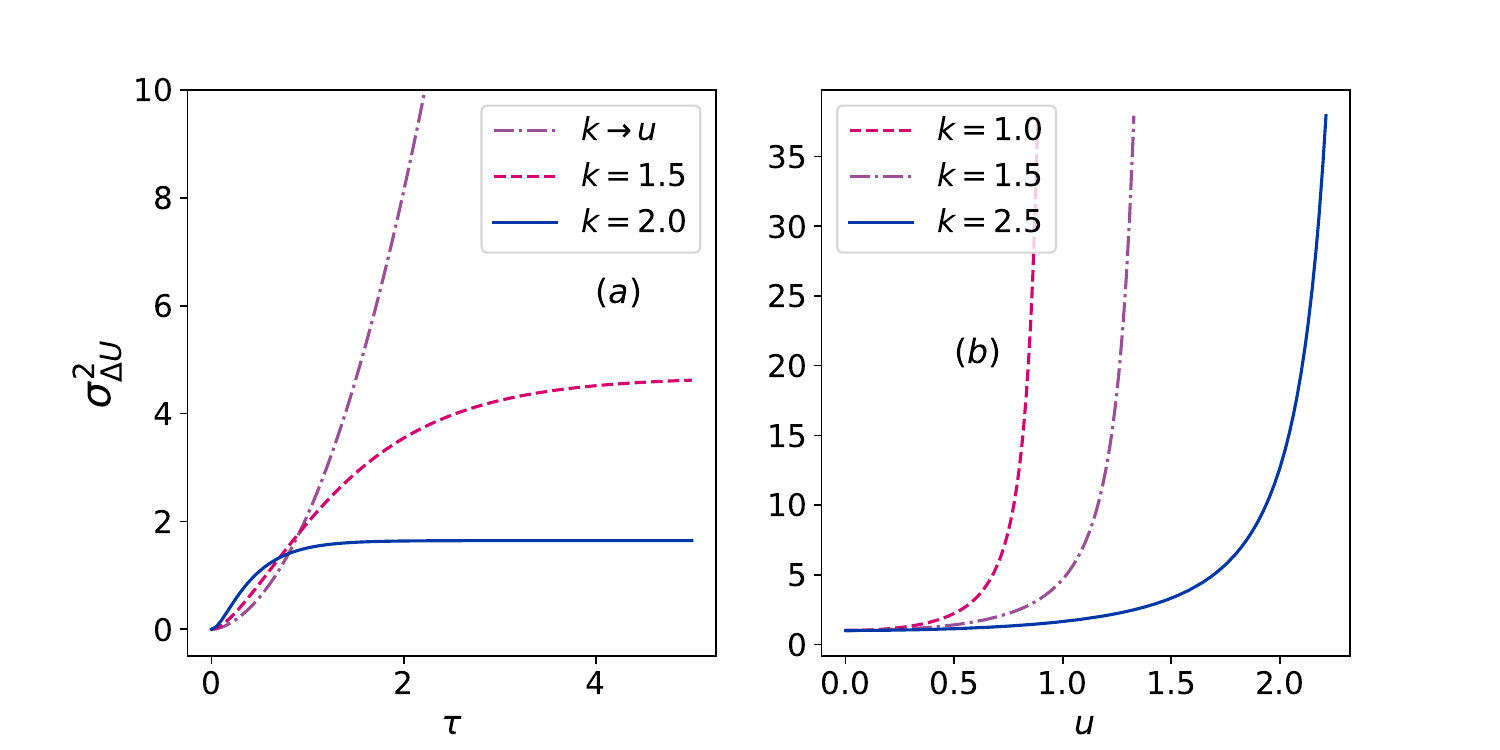}
    \caption{(a) Variance $\sigma^{2}_{\Delta U}$ varying with the time $\tau$, for $k \rightarrow u = 1.0, k = 1.5$ and $k = 2.5$, $\gamma = 1.0$ and $T = 1.0$. For $k \rightarrow u = 1.0$, our results appoint for the divergence of the variance as times varies. For $k \ne u$, our results show the stabilization of $\sigma^{2}_{\Delta U}$ with time. (b) Variation of $\sigma^{2}_{\Delta U}$ with $u$ at the asymptotic limit. Our results appoint the divergence of $\sigma^{2}_{\Delta U}$ with $u \rightarrow k$.}
    \label{variance}
\end{figure}

The next central moment is the variance, and is given by
\begin{equation}
    \sigma^{2}_{\Delta U} = 2 k^2 T^2 \left(\frac{1}{(k+u)^2 \left(f_C(k + u)+1\right)^2}+\frac{1}{(k-u)^2 \left(f_C(k - u)+1\right)^2}\right).\label{varianceharmo}
\end{equation}
This expression is similar to Eq.~\ref{mean_U}. It is the square of the individual terms of Eq.~\ref{mean_U}. Therefore, it is also monotonically increasing in time, as we can see by \ref{variance}a) which shows the temporal evolution of variance for different values of $k$. Notably, as $k$  approaches $u$, we observe a divergence, similar to that of the average. 

Following the divergent behavior of the $k \rightarrow u$ limit of $\mu_{\Delta U}$, the variance does not have an asymptotic limit in this regime. As one can see by
\begin{equation}
    \lim_{k\rightarrow u}\sigma^{2}_{\Delta U}= 2 T^2 \left(\frac{\tau ^2 u^2}{\gamma ^2}+\frac{1}{4 \left(f_C(2u)+1\right)^2}\right)
\end{equation}
This outcome is expected because, in that limit, the particle is in a saddle-point potential, (as depicted in \ref{potential}a). The incapability of the particle to reach equilibrium arises from the non-confining regime. 

As for values of $k>u$, the obtained variance saturates at long times. This behavior is easy to understand by considering the asymptotic limit
\begin{equation}
 \lim_{\tau\rightarrow\infty}\sigma^{2}_{\Delta U} = \frac{k^2 T^2 \left(k^2+u^2\right)}{(k^2-u^2)^2 }.
 \label{asymptoticvariance}
\end{equation}
which shows the divergence for $k\rightarrow u$. Depicted in \ref{potential}a) we can observe that for higher values of $k$, the saturation occurs more rapidly, which is logical since the confinement is stronger.

Analyzing the variation of $\sigma^{2}_{\Delta U}$ with $u$ in the asymptotic limit, Eq.\ref{asymptoticvariance}, in \ref{variance}b), one can observe that $\sigma^{2}_{\Delta U}$ varies positively with $u$, and this happens due to the growth of fluctuations from coupling between positions. Furthermore, by looking at \ref{variance}a), notice that as $k$ increases, the growth of $\sigma^{2}_{\Delta U}$ with $u$ is suppressed. Such behavior can also be observed for very large values of $k$, where the variance in the asymptotic limit becomes constant, that is
\begin{equation}
\lim_{\tau\rightarrow\infty}\sigma^{2}_{\Delta U} \asymp T^2, \;\;\; k\gg 1,
\end{equation}
where $\asymp$ represents the asymptotic equality for larger values of $k$.

The next quantities - skewness, and excess kurtosis - can be expressed analytically, with their analytical forms presented on \ref{appC}. Since the expressions are long and complicated, we adopt a graphical analysis to explore and interpret these statistical measures.
\\

\begin{figure}[h]
    \centering
    \includegraphics[width=1.0\textwidth]{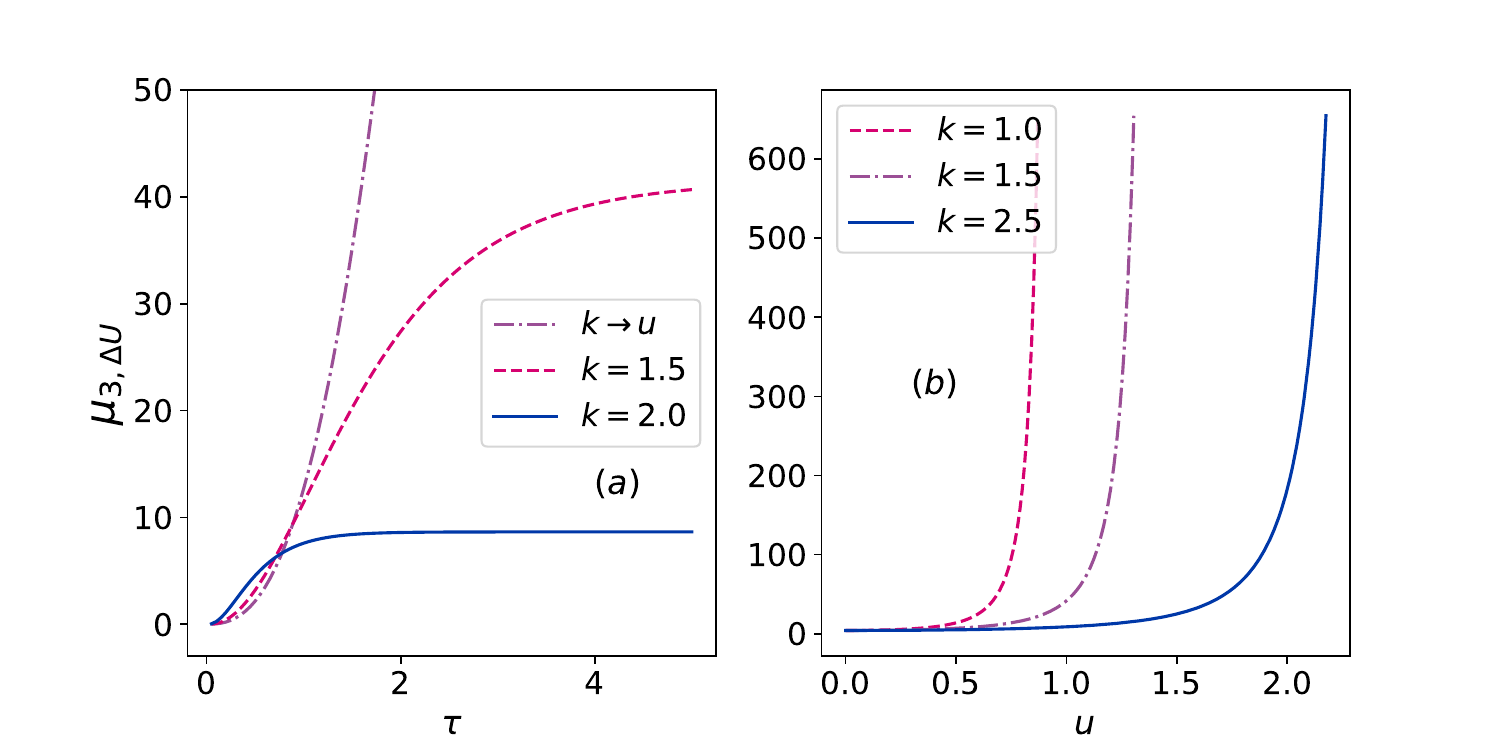}
    \caption{(a) Skewness $\mu_{3,\Delta U}$ varying with the time $\tau$, for $k \rightarrow u = 1.0, k = 1.5$ and $k = 2.5$, $\gamma = 1.0$ and $T = 1.0$. For $k \rightarrow u = 1.0$, our results appoint for the divergence of the variance as times variate, being $\mu_{3,\Delta U}$ positive for all regimes of $k$. For $k \ne u$, our results show the stabilization of $\mu_{3, \Delta U}$ with time. (b) Variation of $\mu_{3, \Delta U}$ with $u$ at the asymptotic limit, for regimes of $k \rightarrow u = 1.0, k = 1.5$ and $k = 2.5$. Similar to $\sigma^{2}_{\Delta U}$, $\mu_{3,\Delta U}$ diverges as $u \rightarrow k$.}
    \label{skewness}
\end{figure}

By means of the skewness, we measure the asymmetry of the distribution. Its formula is written in Eq.\ref{skewness_harm_expression}. Remembering that the initial distribution of positions is a delta centered at the origin, we will only have positive values for the harmonic contribution. Therefore, the skewness will always be positive, growing monotonically over time, as can be seen in \ref{skewness} a). In \ref{skewness} for $k > u$, we observe that the skewness evolves over time, saturating at a certain asymptotic value, which will be
\begin{equation}
   \lim_{\tau \rightarrow\infty} \mu_{3,\Delta U} = \frac{k^4 T^3 \left(4 k^2+7 u^2\right)}{(k^2-u^2)^3}.
\end{equation}
By this formula, we also can see, that as $k$ approaches $u$, we have a positive divergence in the skewness, a consequence of being in the non-confining regime and our skewness not being normalized. Moreover, in \ref{skewness} b), we observe the behavior of $\mu_{3, \Delta U}$ with respect to $u$ for different values of $k$ in the asymptotic time regime. As we increase $u$ we increase the skewness of the harmonic potential, since this increase the fluctuations.

\begin{figure}[h]
    \centering
    \includegraphics[width=1.0\textwidth]{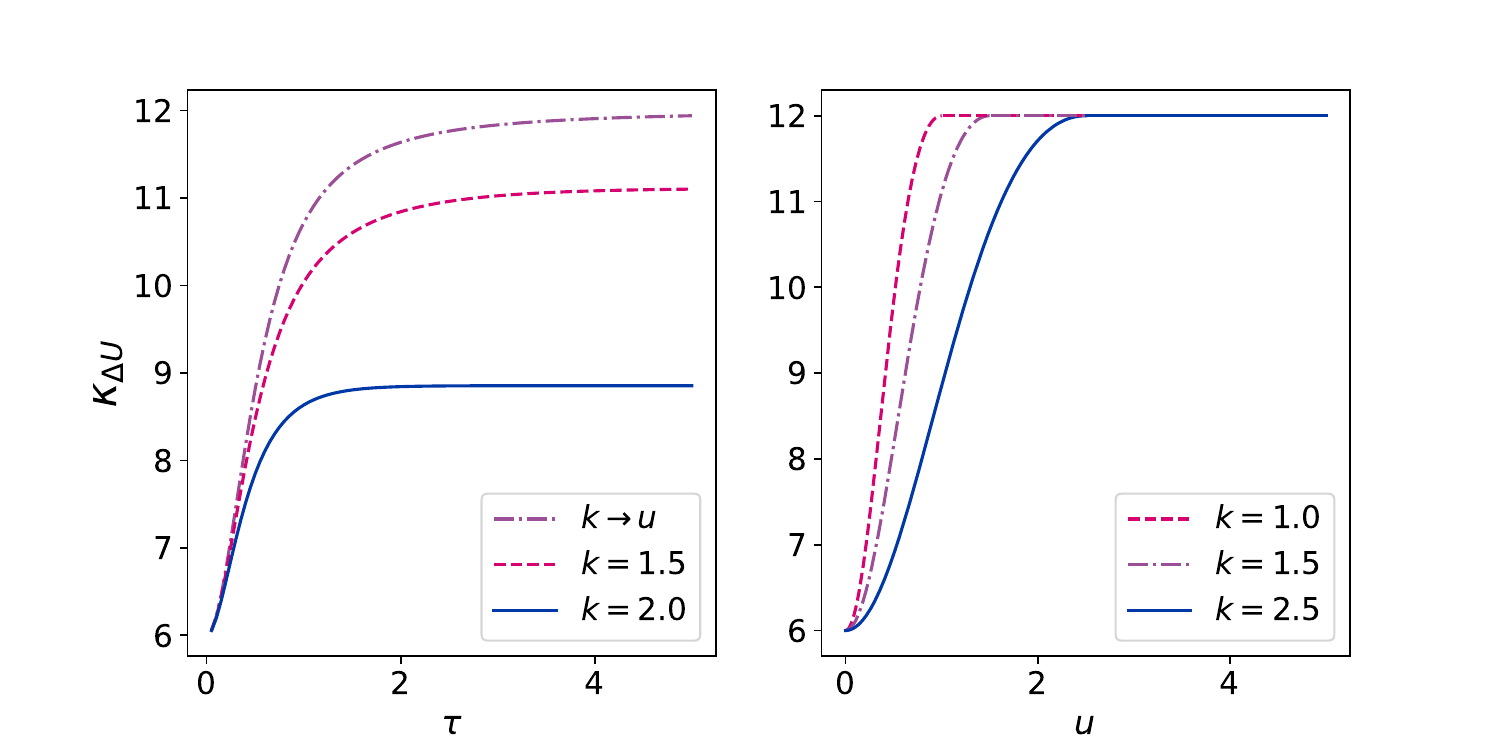}
    \caption{(a) Kurtosis $\kappa_{\Delta U}$ varying with the time $\tau$, for $k \rightarrow u = 1.0, k = 1.5$ and $k = 2.5$, $\gamma = 1.0$ and $T = 1.0$. For $k \rightarrow u = 1.0$. For all $k$ values, our results show the stabilization of $\kappa_{\Delta U}$ with time. (b) Variation of $\kappa_{\Delta U}$ with $u$ at the asymptotic time regime, for values of $k \rightarrow u = 1.0, k = 1.5$ and $k = 2.5$. Our results appoint the monotonic growth of $\kappa_{\Delta U}$ with $u$, that stabilizes as $u \rightarrow k$.}
    \label{curtexc}
\end{figure}

On the other hand, the excess kurtosis measures the behavior of outlier values in the distribution, particularly associated with the tails of the distribution. It is defined as the normalized fourth moment minus 3, as given by equation \ref{curt_harm_expression}. In \ref{curtexc}a), we present the temporal evolution of excess kurtosis, which consistently remains positive for all times, indicating that the distribution is leptokurtic. This means it has a higher probability of rare events occurring compared to a normal distribution \cite{darlington1970kurtosis}. Unlike the other quantities, we observe that kurtosis does not diverge as $k\rightarrow u$. This is because it is a normalized quantity, and the divergence of the fourth central moments cancels with the divergence of the variance. The asymptotic time behavior is
\begin{equation}
    \lim_{\tau\rightarrow\infty}\kappa_{\Delta U} =\begin{cases}
    \frac{24u^2 k^2}{\left(k^2+u^2\right)^2} + 6, \text{ if } k < u,
        \\
        12, \text{ if } k \geq u.
        \end{cases}
\end{equation}
That is, for the asymptotic limit, it stabilizes at $12$ as $u \rightarrow k$, independently of $k$ or $u$. Moreover, similar to the other cases, as $u$ increases, the asymptotic excess kurtosis also increases, as we can see in \ref{curtexc}b), leading to more fluctuations in the tail of the distribution.

\subsection{Distribution}

The distribution is obtained by taking the Fourier transform of the characteristic function, that is
\begin{equation}
    P(\Delta U) = \int \frac{d\lambda}{2\pi}e^{i\lambda \Delta U}Z_{\Delta U}(\lambda),
\end{equation}
however, for the harmonic potential, due the signal of the coefficients, $\alpha_i'$s, we cannot find an analytic expression. 

\begin{figure}[t]
    \centering
    \includegraphics[width=1.0\textwidth]{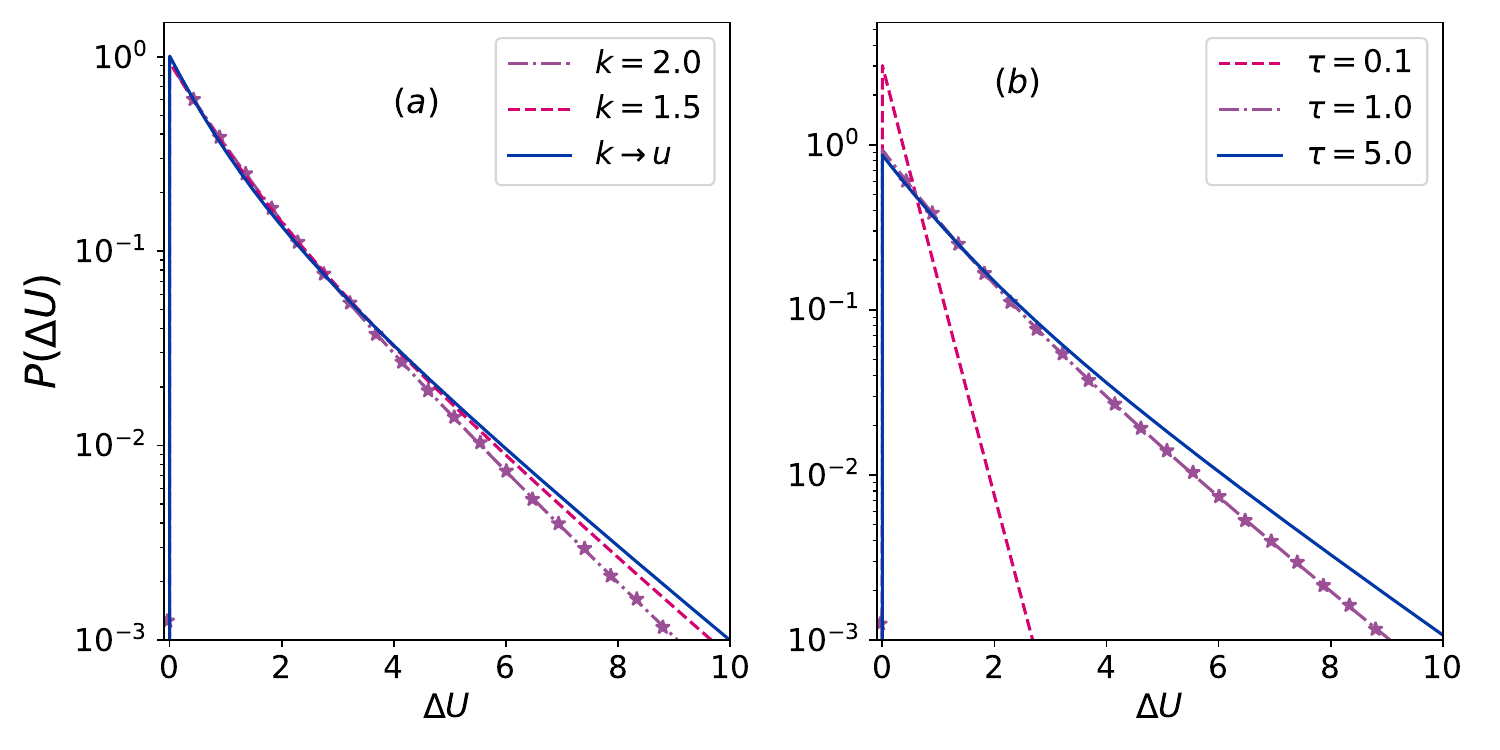}
    \caption{Probability distribution of the harmonic potential $P(\Delta U$) in variation with $\Delta U$, and the purple stars represent the results for simulations of the Langevin equation with $k = 2$ and $\tau = 1.0$. On the left side, the distributions for some values of $k$ are depicted, with $u = 1$, $\gamma = 1$, $T = 1$, and $\tau = 1$. For the harmonic potential, the behavior of $P(\Delta U)$ is similar for $\tau = 1$ for all values of $k$ depicted here, as reported by all calculated moments. The right side depicts $P(\Delta U)$ for some values of $\tau$ and parameters $u = 1$, $\gamma = 1$, $T = 1$, and $k = 2$. For $\tau = 0.1$, the distribution has smaller kurtosis, as the initial position distribution is a delta. The distribution is asymmetric even for asymptotic times as $\Delta U \ge 0$ for all $\{x_{\tau},y_{\tau}\}.$}
    \label{many_harm_dist}
\end{figure}

As the distribution cannot be calculated from the formula in Eq.~\ref{distbessel}, we compute it numerically and compare with numerical simulations of the Langevin equation. \ref{many_harm_dist} shows $P(\Delta U)$ in variation with $\Delta U$ for many regimes of parameters, with the purple stars being the results for the simulations of the Langevin equation. The right side depicts $P(\Delta U)$ for different $k$ values. It admits only positive values as $U = \frac{k}{2}(x_{\tau}^{2} + y_{\tau}^{2}) \ge 0$, as expected from the skewness. The growth of $\kappa_{\Delta U}$ as $k \rightarrow u$ happens in agreement with the results of \ref{curtexc}. In \ref{curtexc}-a, we illustrate $P(\Delta U)$ for various values of $\tau$, whereas in \ref{curtexc}-b, we do the same for distinct values of $u$. Since the particle is initially centered at the origin of the coordinate system, with the position probability distribution being a delta, $P(\Delta U)$ exhibits lower kurtosis for shorter times. This is due to the initial distribution being highly concentrated around $\Delta U = 0$.
 
\section{Fluctuations of the non-confining potential\label{nonconfining}}

Next, we calculate the fluctuations for the non-confining potential $\Delta V_{nc}$ contribution. The variation can be expressed as
\begin{equation}
    \Delta V_{nc} = -u(x_{\tau}y_{\tau} - x_{0}y_{0}).
\end{equation}
As in the previous case, to construct the characteristic function $Z_{nc}(\lambda)$ we only needs the $\{\alpha_{i}\}$ set of coefficients, as showed in the \ref{appB}, represented in Eq.~(\ref{zlambda}). Therefore, for the coefficients we have
\begin{eqnarray}
    \alpha_{1} = 4T^2 u^2,\\
    \alpha_{2} = -2iTu\left((u-k) f_C(k - u) + (k+u)f_C(k + u) + 2u\right),\\
    \alpha_{3} = (k-u) ((k+u) \left(f_C(k + u)+1\right)+f_C(k - u) \left((k+u) f_C(k + u) + u\right)).
\end{eqnarray}
With these coefficients we can construct the characteristic function, as showed in Eq.~\ref{zlambda}, allowing the calculation of moments and distribution.

\subsection{Central Moments}

Following the previous case, we again use the characteristic function to extract the moments of the distribution of $\Delta V_{nc}$ by taking derivatives of $Z_{nc}(\lambda)$ and construct the mean, variance, skewness and excess of kurtosis. As already mentioned, the average is given by the first moment, which is
\begin{equation}
    \langle \Delta V_{nc}\rangle = uT\left(\frac{1}{(k+u)(f_C(k + u) + 1)}-\frac{1}{(k-u) \left(f_C(k - u) +1\right)}\right),
\end{equation}
which is remarkably similar to Eq.~(\ref{mean_U}). But here, we have the coupling $u$ multiplying all the expression, and a difference of two terms, instead of a sum as in  Eq.~(\ref{mean_U}).  When we consider $k > u$, the asymptotic limit gives
\begin{eqnarray}
    \lim_{\tau \rightarrow \infty} \langle \Delta V_{nc}\rangle = -\frac{T u^2}{k^2-u^2}.
\end{eqnarray}
This implies that the asymptotic limit is divergent when $k = u$, similar to the behavior of $\langle \Delta U\rangle$. This is expected, as in the non-confining regime, the particle is not confined to a specific region, and moreover, the mean is negative. We can also derive an expression for the mean in the non-confining regime, but only for a finite transient time. This expression is as follows:
\begin{equation}
    \lim_{k \rightarrow u} \left\langle \Delta V_{nc} \right\rangle = \frac{T}{4}  \left(-\frac{4 \tau  u}{\gamma }-e^{-\frac{4 \tau  u}{\gamma }}+1\right).
\end{equation}
Note that there is also a term with linear dependence with $\tau$, meaning it will negative diverge at $\tau \rightarrow \infty$. 

\begin{figure}[t]
    \centering
    \includegraphics[width=1.0\textwidth]{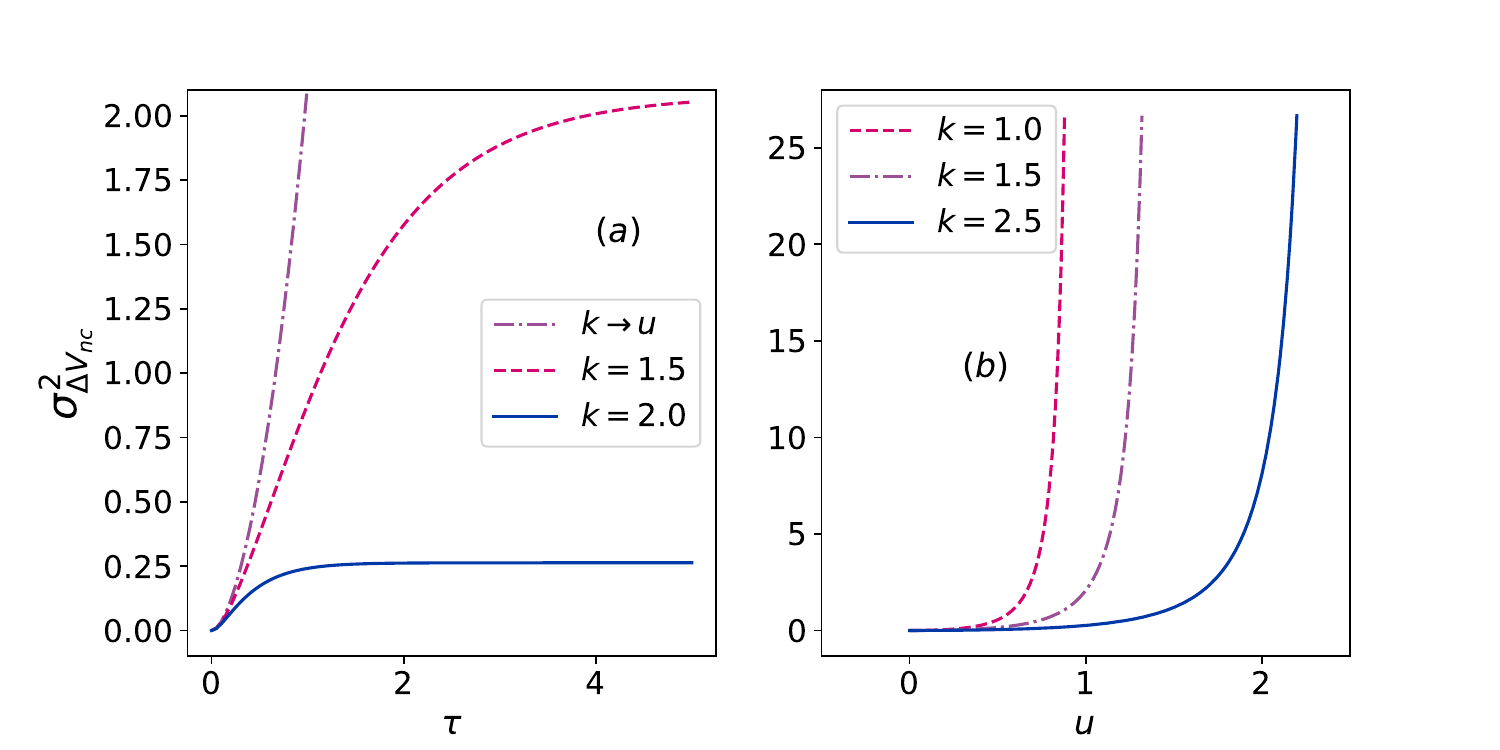}
    \caption{(a) Variance $\sigma^{2}_{\Delta V_{nc}}$ varying with the time $\tau$, for $k \rightarrow u = 1.0, k = 1.5$ and $k = 2.5$, $\gamma = 1.0$ and $T = 1.0$. For $k \rightarrow u = 1.0$, our results appoint for the divergence of the variance with time. For $k \ne u$, our results show the stabilization of $\sigma^{2}_{\Delta U}$ with time. (b) Variation of $\sigma^{2}_{\Delta U}$ with $u$ at the asymptotic limit. Our results appoint the divergence of $\sigma^{2}_{\Delta U}$ as $u \rightarrow k$.}
    \label{variance_uxy}
\end{figure}

The variance  of $\Delta V_{nc}$ will be given by
\begin{equation}
    \sigma_{\Delta V_{nc}}^{2} = 2 T^2 u^2 \left(\frac{1}{(k+u)^2 \left(f_{C}(k+u) + 1\right)^2}+\frac{1}{(k-u)^2 \left(f_{C}(k-u)+1\right)^2}\right).\label{variancenoncon}
\end{equation}
For variance, in the confining regime, we observe a monotonically increasing evolution that saturates at a plateau in the asymptotic time, reducing the rate of evolution as we increase the harmonic coupling $k$. This increased coupling enhances confinement, limiting fluctuations, and consequently decreasing variance. However, in the non-confining regime, when $k = u$, variance asymptotically diverges, as seen in the transient formula
\begin{equation}
   \lim_{k \rightarrow u} \sigma^{2}_{\Delta V_{nc}} = \frac{1}{2} T^2 \left(\frac{4 \tau ^2 u^2}{\gamma ^2}+\frac{1}{\left(f_C(2u)+1\right)^2}\right)
\end{equation}
which diverges for long times $\tau \rightarrow \infty$. Nevertheless, in the confining regime, $k>u$, we have an asymptotic time behavior given by
\begin{equation}
    \lim_{\tau \rightarrow \infty} \sigma^{2}_{\Delta V_{nc}} = \frac{T^2 u^2 \left(k^2+u^2\right)}{(k-u)^2 (k+u)^2}.
\end{equation}
Analyzing the variance in the asymptotic limit, as can be seen in  \ref{variance_uxy}(b), as $u$ increases and the particle becomes unbounded, we notice that the variance diverges, that is (for $u \rightarrow k$, it diverges as in the denominator $k\rightarrow u$ makes an $1/0$ appear). Interesting, the variance of the non confining contribution has the same behavior of the harmonic contribution, we can relate both equations by $\sigma^2_{\Delta V_{nc}}/u=\sigma^2_{\Delta U}/k,$ as one can see by Eq.~\ref{variancenoncon} and \ref{varianceharmo}.

\begin{figure}[t]
    \centering
    \includegraphics[width=1.0\textwidth]{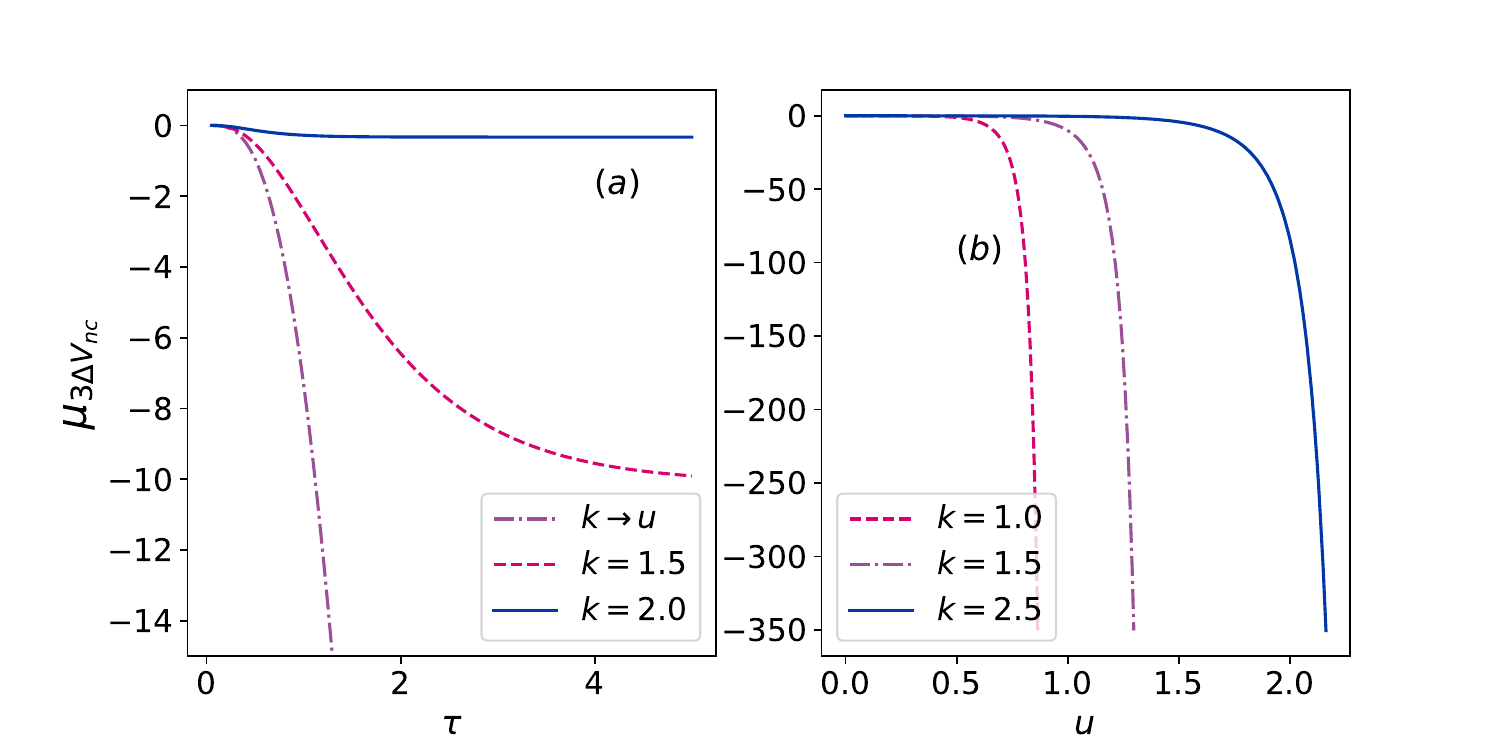}
    \caption{(a) Skewness $\mu_{3, \Delta V_{nc}}$ variating with the time $\tau$, for $k \rightarrow u = 1.0, k = 1.5$ and $k = 2.5$, $\gamma = 1.0$ and $T = 1.0$. For $k \rightarrow u = 1.0$, our results appoint for the  negative divergence of the skewness with time. For $k \ne u$, our results show the stabilization of $\mu_{3, \Delta V_{nc}}$ with time. (b) Variation of $\mu_{3, \Delta V}$ with $u$ at the asymptotic limit. Our results appoint the negative divergence of $\mu_{3, \Delta V_{nc}}$ with $u$.}
    \label{skewness_uxy}
\end{figure}

We proceed to investigate the skewness of the distribution of $\Delta V_{nc}$, given by equation \ref{nc_skew_expression}. Again, because of the initial Dirac delta distribution, the fluctuations are asymmetric. However, different from the previous case, it is negative. Meaning the left area of the distribution is bigger than the positive. In \ref{skewness_uxy}, (a) illustrates the behavior of $\mu_{3}$ on $\tau$ for various values of $k$. Our results show a negative divergence of $\mu_{3}$ as $k$ approaches $u$ in the non-confining regime. \ref{skewness_uxy} (b) depicts the behavior of $\mu_{3}$ varying with $u$ for the asymptotic limit, following the expression 
\begin{equation}
    \lim_{\tau \rightarrow \infty}\mu_{3,\Delta V_{nc}} = -\frac{T^3 u^4 \left(7 k^2+4 u^2\right)}{(k^2-u^2)^3 } .
\end{equation}
Showing that when $k\rightarrow u$ we have a negative divergence as expected in the non-confining regime. Compared with the harmonic case, our skewness in the non-confining contribution contains smaller values. And this will be reflected in the distribution as we will see. Unlike the harmonic case, the non-confining contribution allows us to have positive and negative values in the distribution. 

\begin{figure}[t]
    \centering
    \includegraphics[width=1.0\textwidth]{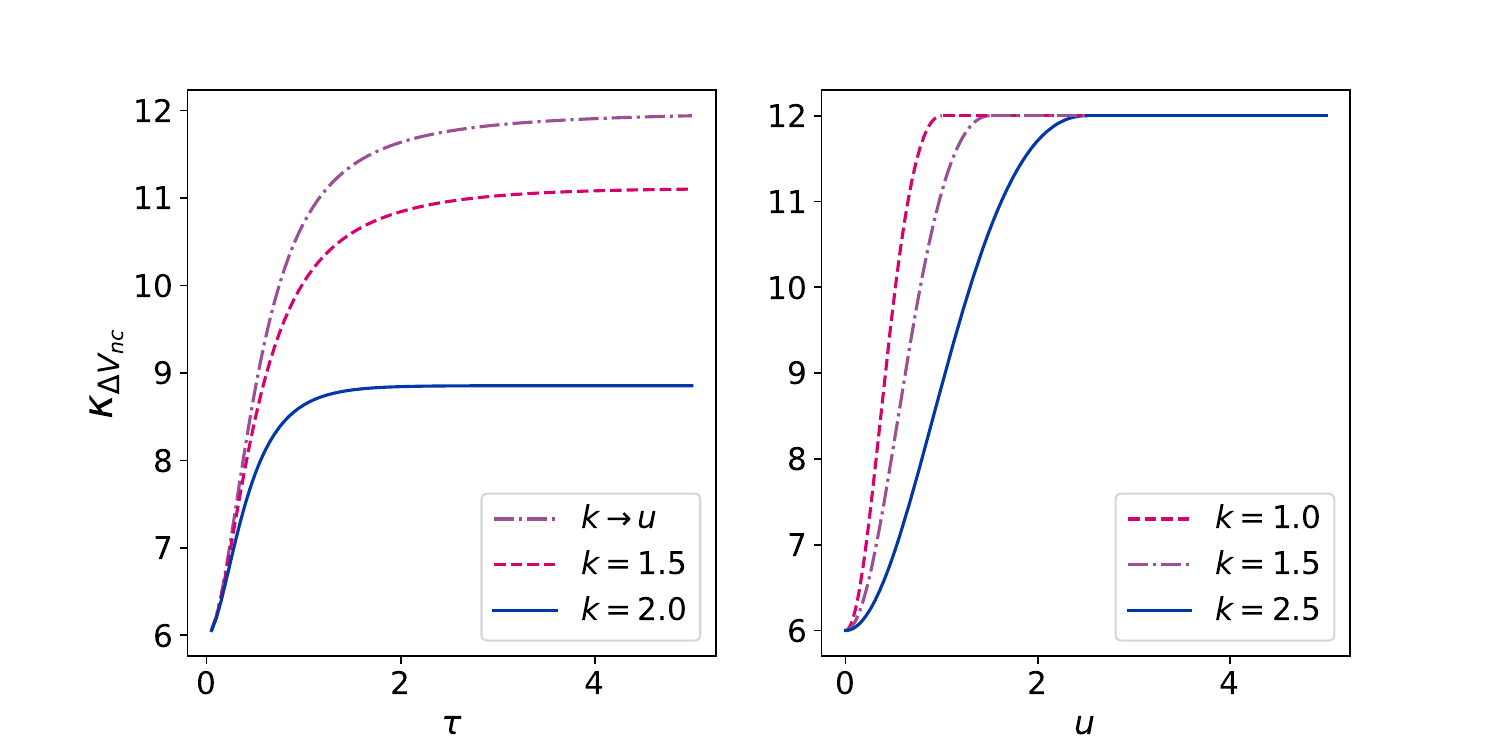}
    \caption{(a) Kurtosis $\kappa_{\Delta V_{nc}}$ variating with the time $\tau$, for $k \rightarrow u = 1.0, k = 1.5$ and $k = 2.5$, $\gamma = 1.0$ and $T = 1.0$. For $k \rightarrow u = 1.0$, our results appoint for the growth and stabilization of $\kappa_{\Delta V_{nc}}$ with $\tau$. (b) Variation of $\kappa_{\Delta V_{nc}}$ with $u$ at the asymptotic limit for $k=1.0, 2.0, 2.5$.}
    \label{kurtosis_uxy}
\end{figure}

The excess kurtosis is equivalent to the harmonic case, as expressed in Eq.~\ref{curt_harm_expression}, and is illustrated in \ref{kurtosis_uxy}. In \ref{kurtosis_uxy}(a), we observe the variation of the excess kurtosis with respect to $\tau$, demonstrating that this distribution is also leptokurtic. And is monotonic increasing in time. Furthermore, it remains finite as $k$ approaches $u$, stabilizing at a specific value in the asymptotic limit as $\tau \rightarrow \infty$. This is a consequence of it being the normalized fourth moment.

In \ref{kurtosis_uxy}(b), we display the variation of the excess kurtosis with $u$ in the asymptotic limit, which is given by
\begin{equation}
\lim_{\tau\rightarrow\infty}\kappa_{\Delta V_{nc}} =\begin{cases}
    \frac{24u^2 k^2}{\left(k^2+u^2\right)^2} + 6, \text{ if } k < u,
        \\
        12, \text{ if } k \geq u.
        \end{cases}
\end{equation}
demonstrating that it stabilizes with the growth of $u$ as $u \rightarrow k$.
Having analyzed the most significant central moments, it becomes evident that the even moments (kurtosis and variance) are closely related to the harmonic case, revealing similarities between these two distinct quantities.

\subsection{Distribution}
Since for the non-confining contribution, $\Delta V_{nc}$, the coefficient $\alpha_1$ is positive, we are able to find a analytic expression for the distribution (See Eq.~(\ref{distbessel}) in the \ref{appB}). It is given by
\begin{eqnarray}
    P(\Delta V_{nc}) = \frac{1}{2 \pi }\exp \left(-\frac{\Delta V_{nc} \left((u-k) f_C(k-u)+(k+u) f_C(k+u)+2 u\right)}{4 T u}\right)\nonumber \\  \times K_0\left(\frac{1}{4} |\Delta V_{nc}|  \frac{\left((k-u) f_C(k-u)+(k+u) f_C(k+u)+2 k\right)}{T u}\right)\nonumber \\ \times \sqrt{\frac{(k-u) (k+u) \left(f_C(k-u)+1\right) \left(f_C(k+u)+1\right)}{T^2 u^2}}.
   \label{dist_bessel}
\end{eqnarray}

\begin{figure}[t]
    \centering
    \includegraphics[width=1.1\textwidth]{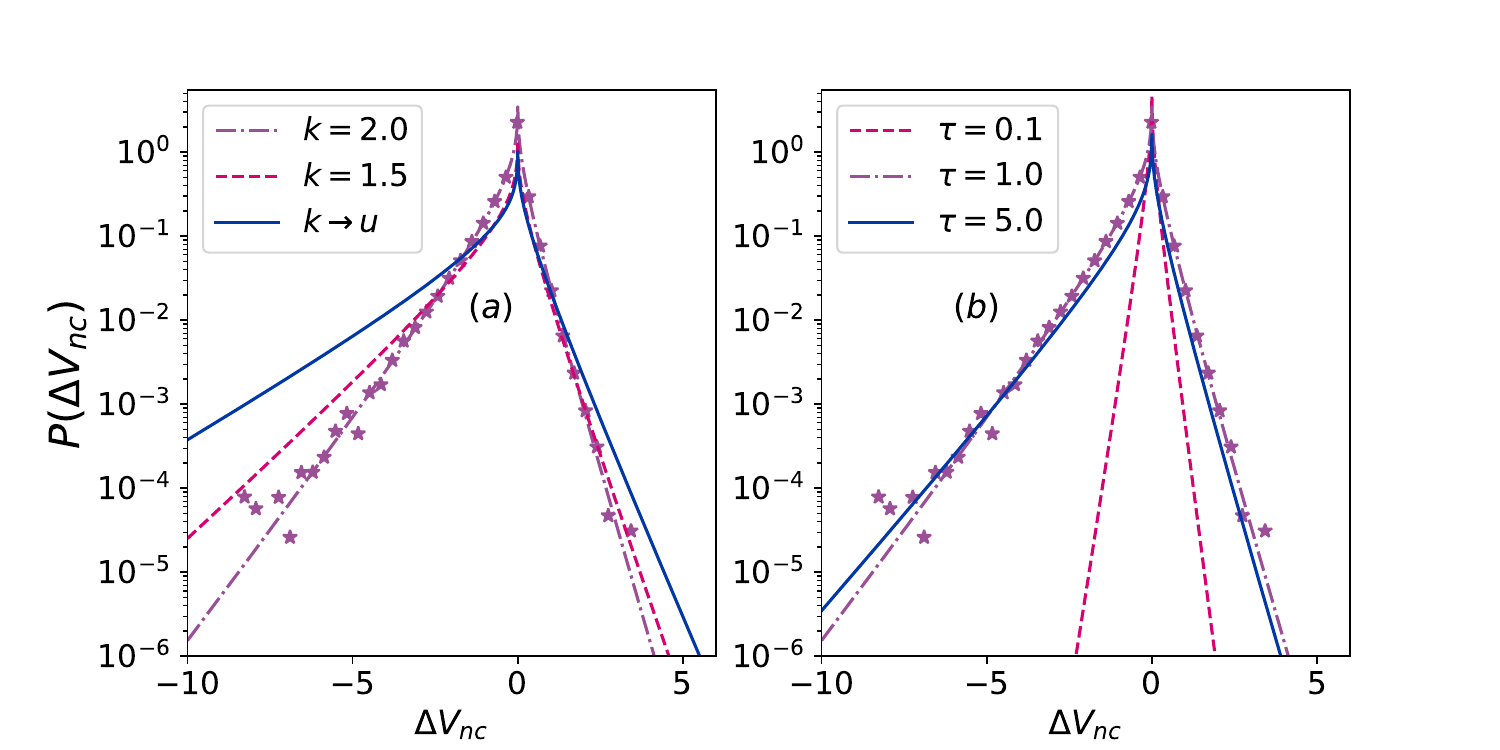}
    \caption{Probability distribution of the non-confining potential $P(\Delta V_{nc})$, varying with $\Delta V_{nc}$, as expressed in Eq.~(\ref{dist_bessel}). The purple stars represent the results for the simulations of the Langevin equation. On the left side, distributions for some values of $k$ are depicted, with $u = 1$, $\tau = 1$, $\gamma = 1$, and $T = 1$. Due to the singularity on the Bessel function for $\Delta V_{nc} = 0$, the probability appears to be infinity at this region. This is, however, not the physical reality of this system, that can be seen from the purple stars for $k = 2.0$. The right side depicts $P(\Delta V_{nc})$ for many regimes of $\tau$ and $k = 2.0$, $u = 1.0$, $\gamma = 1.0$, and $T = 1$. Due to the initial delta distribution, $P(\Delta V_{nc})$ for $\tau = 0.1$ is highly concentrated at $\Delta V_{nc} = 0$. This behavior changes as $\tau$ grows. However, for our analytical expressions are given in terms of $P(\Delta V_{nc}) \propto K_{0}(\Delta V_{nc})$, there is also a divergence at $\Delta V_{nc} = 0$.}
    \label{Many_nc_dist}
\end{figure} 

\ref{Many_nc_dist} depicts the probability distribution $P\left(\Delta V_{nc}\right)$ varying with $\Delta V_{nc}$ for many regimes of parameters given by eq.~(\ref{dist_bessel}), with the purple stars being the results for the simulations of the Langevin equation. The right side depicts $P(\Delta V_{nc})$ for many $k$ values. Being the distribution proportional to $K_{0}\left(|\Delta V_{nc}|\right)$, it will diverge for $\Delta V_{nc} = 0$, which is usual for Bessel type distributions \cite{chatterjee2010exact,paraguassu2023heat}, for instance another distribution having this behavior is a  Dirac delta distribution. This divergence does not affect the normalization of the distribution.

Aside from this point, the tails and width of $P\left(\Delta V_{nc}\right)$ grow as $k \rightarrow u$, in agreement with the behavior of the excess of kurtosis and the variance. This left side of the distribution increases, while the right side decreases, as expected from the skewness.  Moreover, in \ref{Many_nc_dist} b), one can observe $P(\Delta V_{nc})$ in variation with $\Delta V_{nc}$ for different values of $\tau$. The behavior is in agreement with the predictions from the central moments.

In the non-confining regime, the distribution obeys a relation analogous to the Fluctuation theorem \cite{seifert2012stochastic}. By
calculating the log ratio between the distribution of $\Delta V_{nc}$ and its reverse process, given by distribution $P(-\Delta V_{nc})$, one gets
\begin{equation}
    \log\left(\frac{P(\Delta V_{nc})}{P(-\Delta V_{nc})}\right) = -\frac{\Delta V_{nc} \left((u-k) f_C(k-u) + (k+u)f_C(k+u) + 2 u\right)}{2 T u}.
\end{equation}
Now considering for the non-confining regime $k \rightarrow u$, it becomes
\begin{equation}
    \lim_{k\rightarrow u} \log\left(\frac{P(\Delta V_{nc})}{P(-\Delta V_{nc})}\right) = -\Delta V_{nc} \left(-\frac{\gamma}{2\tau Tu} + \frac{\coth \left(\frac{2 \tau  u}{\gamma }\right) + 1}{T}\right),
\end{equation}
and for a asymptotic time we have
\begin{equation}
    \lim_{\tau \rightarrow \infty}\lim_{k\rightarrow u} \log\left(\frac{P(\Delta V_{nc})}{P(-\Delta V_{nc})}\right) = -\frac{2\Delta V_{nc}}{T}.
    \label{fluctuation_theorem}
\end{equation}
Which is a Ratio of Irreversibility (RI) \cite{evans2002fluctuation}.  This RI is a mathematical consequence of the expression for $P(\Delta V_{nc})$ that admits proportional probabilities for both $\Delta V_{nc} > 0$ and $\Delta V_{nc} < 0$. Even though it is similar to Crooks fluctuation theorem \cite{crooks1999entropy}, as far as we know, it does not necessarily has any physical relation to it, being a mathematical consequence of the non-confining contribution and the behavior of the dynamics in the non-confining regime.

\section{Fluctuations of the full potential\label{fullpotential}}

Finally, having investigated the two contributions of the total potential. We proceed to investigate the full variation of potential energy, that can be expressed as:
\begin{equation}
    \Delta V = \Delta U + \Delta V_{nc} = \frac{k}{2}(x_{\tau}^2 + y_{\tau}^2 - x_{0}^2 - y_{0}^2) - u(x_{\tau}y_{\tau} - x_{0}y_{0}).
\end{equation}
Again, as introduced in the previous sections, we only need $\{\alpha_{i}\}$ set of coefficients, which comes from the fact that the potential is quadratic and the positions are Gaussian variables. 

Thus, the coefficients of the characteristic function, Eq.~\ref{zlambda}, are given by:
\begin{eqnarray}
    \alpha_{1} = -4T^2(k^2 - u^2),\\
    \alpha_{2} = 2iT(k^2 - u^2)(2 + f_C(k - u) + f_C(k + u)),\\
    \alpha_{3} = (k^2-u^2) \left(f_C(k - u) + 1\right) \left(f_C(k + u) + 1\right).
\end{eqnarray}
Having the characteristic function, we proceed to calculate the central moments.

\subsection{Central Moments}

\begin{figure}[h]
    \centering
    \includegraphics[width=1.0\textwidth,left]{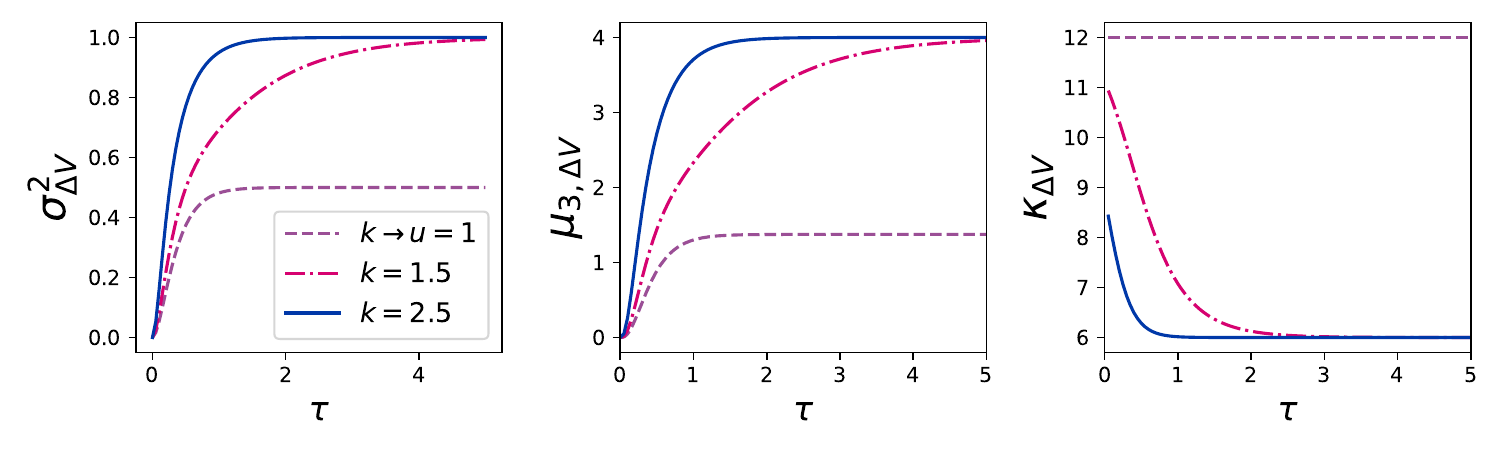}
    \caption{Variation of Central Moments of the total potential distribution with time $\tau$. These computations were carried out for three distinct regimes of $k$: $k = 1$, $k = 1.5$, and $k = 2.5$. For all regimes, parameters $u = 1$, $\gamma = 1.0$, and $T = 1.0$ were adopted. a) Evolution of the variance of the total distribution $\sigma^{2}_{\Delta V}$ with the time $\tau$. Our results appoint the stabilization of $\sigma^{2}_{\Delta V}$ with time, in accordance with the results for asymptotic time. b)  Skewness of the total distribution $\mu_{3,\Delta V}$ variating with the time $\tau$. For all $k$ regimes, our results appoint for the stabilization of $\mu_{3,\Delta V}$, in accordance with asymptotic time results. c) Kurtosis of the total distribution $\kappa_{\Delta V}$ variating with the time $\tau$. For all $k \ne u$ regimes, there is a decreasing and stabilization of $\kappa_{\Delta V}$ as $\tau$ variates. For $k \rightarrow u$, our results point to a constant $\kappa_{\Delta V}$ with time.}
    \label{variance_fulldist}
\end{figure}

The mean is given by
\begin{equation}
    \langle \Delta V\rangle = T\left(\frac{1}{1 + f_C(k+u)}+\frac{1}{1 + f_C(k - u)}\right),
\end{equation}
which comes from the summation $\langle \Delta V\rangle = \langle \Delta U + \Delta V_{nc}\rangle = \langle \Delta U\rangle + \langle \Delta V_{nc}\rangle$. Note that the dependence on the coupling constant, only appear in the function $f_C$. Different from the previous case, there is no divergence, even in the non-confining regime in the asymptotic time limit, that is
\begin{equation}
    \lim_{\tau \rightarrow \infty}\langle \Delta V\rangle = \begin{cases}
        T \text{ if } k > u,
        \\
        \frac{T}{2} \text{ if } k = u.
        \end{cases}
\end{equation}
Note that the mean becomes proportional to the temperature, and a lower temperature for the non-confining regime. It is similar to the equipartition theorem \cite{jackson2000equilibrium}, meaning that the non-confining regime has lower energy than the confining case.

The variance of the full potential $\Delta V$ is analogous, and can be expressed as
\begin{equation}
    \sigma^{2}_{\Delta V} = 2 T^2 \left(\frac{1}{\left(f_C(k+u) + 1\right)^2}+\frac{1}{\left(f_C(k - u) + 1\right)^2}\right).
\end{equation}
For values $k \geq u$, similarly to the cases of $\Delta U$ and $\Delta V_{nc}$, the variance of the total energy saturates at long times, even for values of $k$ in the limit $k \rightarrow u$. This can be understood as, in the limit of asymptotic time, one gets
\begin{equation}
\lim_{\tau \rightarrow \infty} \sigma^{2}_{\Delta V} = \begin{cases}
        T^2 \text{ if } k > u,
        \\
        \frac{T^2}{2} \text{ if } k = u.
        \end{cases}
\end{equation}
without any dependency on $k$ and $u$. As expected since we are in the asymptotic time. Similar to the mean, we also observe a lower value in the non-confining regime. In \ref{variance_fulldist} (a), the temporal evolution of the variance of $\Delta V$ for different values of $k$ is depicted. It's evident that, apart from the stabilization of $\sigma^{2}_{\Delta V}$ in the asymptotic time regime, the magnitude of $\sigma^{2}$ varies with changes in $k.$ Notably, as $k$ approaches $u$, the saturation occurs at a lower value compared to the confining case, where a higher $k$ results in a faster evolution towards the asymptotic time.

As for the skewness we have a more compact formula compared with the previous cases, we have
\begin{equation}
    \mu_{3,\Delta V} = T^3 \left(\frac{5 \left(f_C(k-u)+f_C(k+u)+2\right)}{\left(f_C(k-u)+1\right)^2 \left(f_C(k+u)+1\right)^2}+ \frac{11}{\left(f_C(k+u)+1\right)^3}+\frac{11}{\left(f_C(k-u)+1\right)^3}\right).
\end{equation}
where in \ref{variance_fulldist} b) depicts the temporal evolution of the skewness $\mu_{3, \Delta V}$, for different values of $k$. Following a similar tendency to the variance, the skewness saturates faster with $k$. And for the asymptotic limit, we have
\begin{equation}
\lim_{\tau \rightarrow \infty} \\\mu_{3,\Delta V} = \begin{cases}
        4T^3 \text{ if } k > u,
        \\
        \frac{11\,T^3}{8} \text{ if } k= u.
        \end{cases}
\end{equation}
We again, do not have a divergence in the non-confining regime. What's happens is that the for $k=u$ the skewness becomes
\begin{equation}
  \lim_{k\rightarrow u} \mu_{3,\Delta V} = \frac{11\, T^3}{\left(f_C(2u)+1\right)^3},
\end{equation}
which does not diverge for $\tau \rightarrow\infty$.

Now, for the excess of kurtosis, we have a long expression, written in Eq.~\ref{tot_kurtosis_expression}. 
\ref{variance_fulldist} c) depicts the temporal evolution of the excess of kurtosis for different values of $k$ with $u$ set to one. we see that, for $k < u $, it decreases with $\tau$, saturating close to $k=u$ for stronger values of $k$. At the non-confining regime, the kurtosis is constant for all times. Its asymptotic limit, is given by
\begin{equation}
\lim_{\tau \rightarrow \infty} \kappa_{\Delta V} = \begin{cases}
        6 \text{ if } k > u,
        \\
        12 \text{ if } k = u.
        \end{cases}
\end{equation}
This demonstrates that the kurtosis is constant and positive for both regimes. Thus, Leptokurtic. Like the previous moments, it exhibits no dependence on temperature. It's worth noting that this is the case because we are dealing with the normalized fourth central moment, which is dimensionless.

\subsection{Distribution}
In a similar fashion to the harmonic case, it is not possible to obtain the distribution from equation \ref{distbessel}. We then proceed to determine numerically the distribution, and compare to the numerical simulations of the Langevin equation.

\begin{figure}[t]
    \centering
    \includegraphics[width=1.0\textwidth]{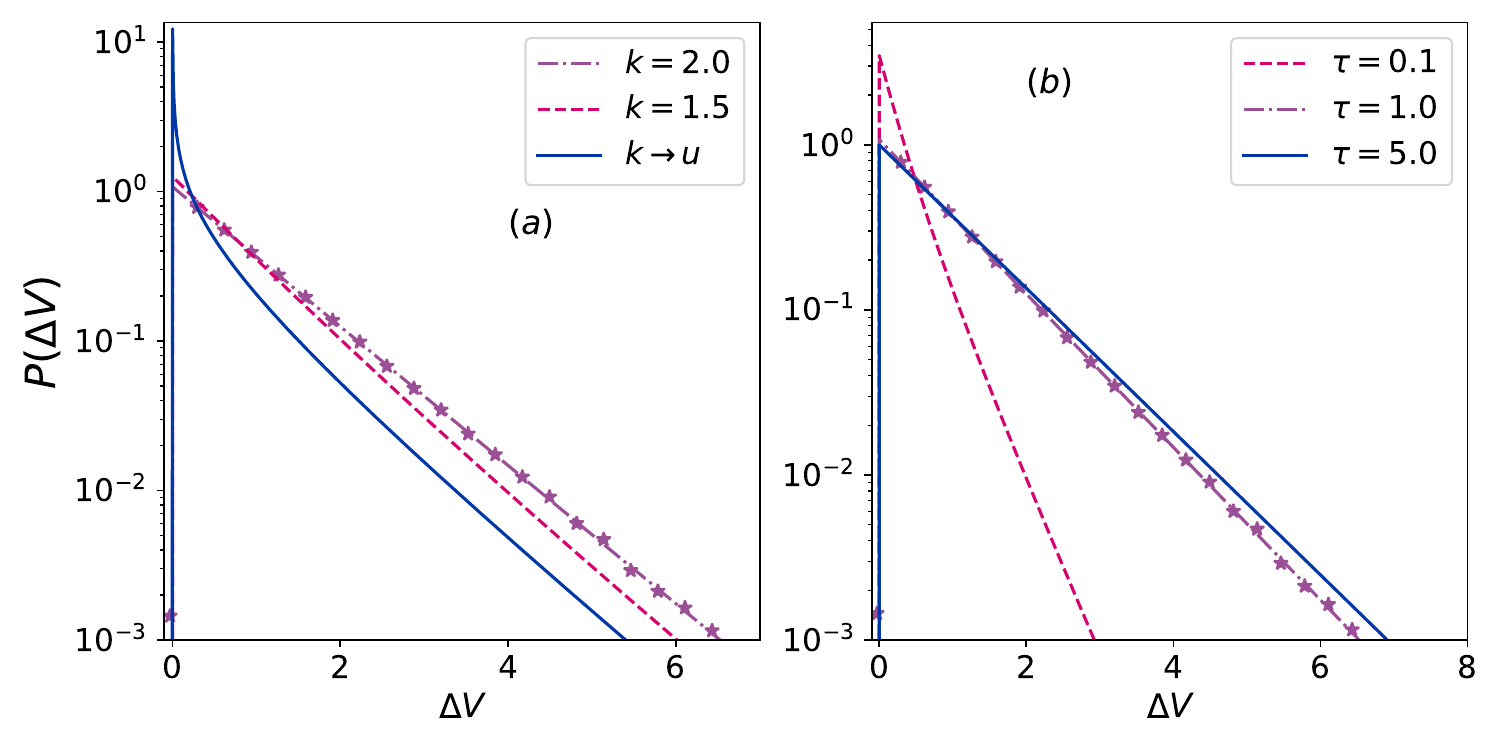}
    \caption{Probability distribution of the total potential $P(\Delta V)$, is varying with $\Delta V$, and the purple stars representing our results for the simulations of Langevin equation. On the left side, distributions for some values of $k$ are depicted, with $u = 1$, $\tau = 1$, $\gamma = 1$, and $T = 1$. The right side depicts $P(\Delta V)$ for many regimes of $\tau$ and $k = 2.0$, $u = 1.0$, $\gamma = 1.0$, and $T = 1$. Due to the initial delta distribution, $P(\Delta V)$ for $\tau = 0.1$ is highly concentrated at $\Delta V = 0$. This behavior changes as $\tau$ grows.}
    \label{Many_total_dist}
\end{figure} 

\ref{Many_total_dist} a) depicts the probability distribution $P(\Delta V)$ in variation of $\Delta V$ for different values of $k$, with the purple stars being the results for the simulations of the Langevin equation. All distributions are completely asymmetric due to $\Delta V \ge 0$ for all values of $x_{\tau}$ and $y_{\tau}$, even in the non-confining regime $k=u$. 

The time evolution of the distribution is showed in \ref{Many_total_dist} b), $P(\Delta V)$ for many $k$ values. It is in agreement with the expected behavior of the central moments, as we see that the distribution becomes broader and skew as predicted by the variance and kurtosis. Besides that, the distribution tail increases compared with the initial time value, as expected from the excess of kurtosis.

While for the non-confining potential $V_{nc}$ our results show the existence of a RI, reported in Eq.~\ref{fluctuation_theorem}, for the full potential the scenario is different. As one can see clearly from the fact that the distribution admits only positives values. Nevertheless, we can discuss what is happening.  For instance, let us consider the full potential difference
\begin{equation}
    \Delta V = \frac{k}{2} (x_{\tau}^{2} + y_{\tau}^{2} - x_{0}^{2} - y_{0}^{2}) - u\,(x_{\tau}y_{\tau} - x_{0}y_{0}).
\end{equation}
Considering the particle at the origin of the coordinates system (in a delta distribution), meaning $x_{0} = y_{0} = 0$, one has
\begin{equation}
    \Delta V = \frac{k}{2} (x_{\tau}^{2} + y_{\tau}^{2}) - u\,x_{\tau}\,y_{\tau}.
\end{equation}
For the RI, one consider the non-confining regime of the potential $k \rightarrow u$. That is, for any pair $(x_{\tau}, y_{\tau})$, one gets
\begin{equation}
    \lim_{k \rightarrow u}\Delta V = \frac{u}{2}(x_{\tau} - y_{\tau})^{2} \ge 0.
\end{equation}
From this result, we must know that the system does not admit regimes of $\Delta V < 0$. That means that the probability $\lim_{k \rightarrow u}P(\Delta V) = 0$ for $\Delta V < 0$. Then, for the fluctuation theorem one has
\begin{equation}
    \lim_{\tau \rightarrow \infty}\lim_{k\rightarrow u} \log \left(\frac{P(\Delta V)}{P(-\Delta V)}\right) \rightarrow 0 \mbox{ or }\infty.
\end{equation}
The same discussion applies to the harmonic case as well.

\section{Discussion and conclusion\label{Discussion}}

We investigated the fluctuations of a quadratic potential associated to a Brownian particle in a parabolic trap, which exhibits two regimes: confining and non-confining. These fluctuations are inherent to the system, that consists of a Brownian particle in two dimensions, with motion starting at the origin. 

We divided the total potential into two contributions: a harmonic one, which is confining, and another non-confining one. We explored the fluctuations of each contribution in both regimes, confining and non-confining. We found that the even central moments (variance and kurtosis) are qualitatively similar, despite the odd moments (mean and skewness) having opposite signs. In the non-confining regime, except for the excess kurtosis, all statistical quantities are observed to diverge in the asymptotic time limit. This is a characteristic of the system, since the position distribution becomes unbounded in this regime, approaching a zero distribution in the asymptotic time limit.

For the total potential, the behavior is different; all quantities remain finite in any regime in the asymptotic limit. Additionally, we observe the emergence of the equipartition of energy theorem, as the potential becomes $\Delta V = N T/2$, with $N = 1$ for the non-confining regime ($k=u$) and $N=2$ for the confining regime. We can interpret this as follows: In the confining regime, $k>u$, we have two potentials, the harmonic one, $\Delta U$, and the non-confining contribution, $\Delta V_{nc}$, resulting in a greater degree of freedom compared to the non-confining regime, where the potential becomes a single term, $\Delta V = k(x-y)^2/2$, contributing to only one degree of freedom.

In addition to the moments, we also computed the distributions and compared the results with numerical simulations, finding agreement. Due to the presence of negative coefficients in the characteristic function, obtaining an analytical expression for the harmonic contribution and the total potential is not feasible. In both cases, we have completely positive distributions due to the initial Dirac delta condition for positions.

However, for the non-confining contribution, the result is analytical and provides an asymmetric Bessel-type distribution, where the asymmetry arises from the exponential multiplied by the Bessel function. Unlike the other cases, this distribution allows for both positive and negative values. This enables the analysis of the irreversibility ratio, where we find an expression similar to Crooks' fluctuation theorem~\cite{Crooks2000}. It's essential to emphasize that this similarity is purely a mathematical consequence of the form of this potential.

In this article, we focused on the potential energy fluctuations over a time interval $t\in[0,\tau]$ and its individual contributions. It is worth noting that the behavior discussed here can be used to interpret heat, which, for the studied system, differs from the potential difference due to kinetic energy, i.e., $Q = \Delta K + \Delta V$  (even in the overdamped regime, as pointed out in \cite{paraguassu2022effects}). Therefore, our work not only investigates potential fluctuations in different regimes but could also contribute to the understanding of another quantities in stochastic thermodynamics.

Furthermore, possible generalizations of this work are feasible. A situation of interest is to consider different thermal baths, each coupled to a degree of freedom, leading to a non-equilibrium steady state (NESS), as usually discussed on the context of Brownian Gyrators.

\section*{Acknowledgments}
This work is supported by the Brazilian agencies CNPq, CAPES, and FAPERJ. This study was financed in part by Coordena\c{c}\~ao de Aperfei\c{c}oamento de Pessoal de N\'{ı}vel Superior—Brasil (CAPES)—Finance Code 001.

\appendix
\section{Path Integrals Calculation}\label{pathcalculations}
In the action formula, the conditional probability is given by
\begin{equation}
    P[x,y,t|x_0,y_0,0] = \int \mathcal{D}x\int \mathcal{D}y 
    \exp\left(-\frac{1}{4\gamma T}S[x,y]\right).
\end{equation}
As the action is quadratic, therefore we just need to calculate the extremized action, and the result will be
\begin{equation}
    P[x,y,t|x_0,y_0,0] \sim 
    \exp\left(-\frac{1}{4\gamma T}S_{d}[x,y]\right).
\end{equation}
The $S_{d}$ action represents the action evaluated at the solution of the Euler-Lagrange equations for the space coordinates, that gives the deterministic path. It is given by:
\begin{eqnarray}
(u^2+k^2)y(t)=2 u x(t)+y''(t)\\
(u^2+k^2) x(t)=2 u y(t)+x''(t).
\end{eqnarray}
Solving both equations with boundary conditions $x(0) = x_{0}$, $y(0) = y_{0}$, $x(\tau) = x_{\tau}$ and $y(\tau) = y_{\tau}$, one gets the solutions \begin{eqnarray}
     x_{c}(t) =\frac{1}{4} e^{\frac{2 k \tau }{\gamma }} \left(\coth \left(\frac{\tau  (k-u)}{\gamma }\right)-1\right) \left(\coth \left(\frac{\tau  (k+u)}{\gamma }\right)-1\right)\nonumber \\ 
   \times\Bigg\{(x_{0}+y_{0}) \cosh \left(\frac{k t-2 k \tau -t u}{\gamma }\right)+(x_{0}-y_{0}) \left(\cosh \left(\frac{t (k+u)-2 k \tau }{\gamma }\right)-\cosh
   \left(\frac{t (k+u)-2 \tau  u}{\gamma }\right)\right)\nonumber \\  -(x_{0}+y_{0}) \cosh \left(\frac{k t-t u+2 \tau  u}{\gamma }\right)+2 x_{\tau} \cosh \left(\frac{k
   (t+\tau )}{\gamma }\right) \cosh \left(\frac{u (t-\tau )}{\gamma }\right)\nonumber\\  -2 x_{\tau} \cosh \left(\frac{k (t-\tau )}{\gamma }\right) \cosh \left(\frac{u (t+\tau
   )}{\gamma }\right)-2 y_{\tau} \sinh \left(\frac{k (t+\tau )}{\gamma }\right) \sinh \left(\frac{u (t-\tau )}{\gamma }\right)\nonumber \\  +2 y_{\tau} \sinh \left(\frac{k
   (t-\tau )}{\gamma }\right) \sinh \left(\frac{u (t+\tau )}{\gamma }\right)\Bigg\},\\ 
     y_{c}(t) =\frac{1}{4} e^{\frac{2 k \tau }{\gamma }} \left(\coth \left(\frac{\tau  (k-u)}{\gamma }\right)-1\right) \left(\coth \left(\frac{\tau  (k+u)}{\gamma }\right)-1\right)\nonumber\\ \times\Bigg\{(x_{0}+y_{0}) \cosh \left(\frac{k t-2 k \tau -t u}{\gamma }\right)-(x_{0}-y_{0}) \left(\cosh \left(\frac{t (k+u)-2 k \tau }{\gamma }\right)-\cosh\left(\frac{t (k+u)-2 \tau  u}{\gamma }\right)\right)\nonumber \\  -(x_{0}+y_{0}) \cosh \left(\frac{k t-t u+2 \tau  u}{\gamma }\right)-2 x_{\tau} \sinh \left(\frac{k(t+\tau )}{\gamma }\right) \sinh \left(\frac{u (t-\tau )}{\gamma }\right)\nonumber \\  +2 x_{\tau} \sinh \left(\frac{k (t-\tau )}{\gamma }\right) \sinh \left(\frac{u (t+\tau)}{\gamma }\right)+2 y_{\tau} \cosh \left(\frac{k (t+\tau )}{\gamma }\right) \cosh \left(\frac{u (t-\tau )}{\gamma }\right)\nonumber \\  -2 y_{\tau} \cosh \left(\frac{k(t-\tau )}{\gamma }\right) \cosh \left(\frac{u (t+\tau )}{\gamma }\right)\Bigg\}.
\end{eqnarray}

Adopting the delta initial condition $x_{0} = \delta(x_{0})$ and $y_{0} = \delta(y_{0})$, the extremized path reduces to
\begin{eqnarray}
     x_{c}(t) =\frac{2 \left(\sinh \left(\frac{k t}{\gamma }\right) \cosh \left(\frac{t u}{\gamma }\right) \left(x_{\tau} \sinh \left(\frac{k \tau }{\gamma }\right) \cosh
   \left(\frac{\tau  u}{\gamma }\right)+y_{\tau} \cosh \left(\frac{k \tau }{\gamma }\right) \sinh \left(\frac{\tau  u}{\gamma }\right)\right)\right)}{\cosh \left(\frac{2 k \tau }{\gamma
   }\right)-\cosh \left(\frac{2 \tau  u}{\gamma }\right)}\nonumber\\ -\frac{\left(2\cosh \left(\frac{k
   t}{\gamma }\right) \sinh \left(\frac{t u}{\gamma }\right) \left(x_{\tau} \cosh \left(\frac{k \tau }{\gamma }\right) \sinh \left(\frac{\tau  u}{\gamma
   }\right)+y_{\tau} \sinh \left(\frac{k \tau }{\gamma }\right) \cosh \left(\frac{\tau  u}{\gamma }\right)\right)\right)}{\cosh \left(\frac{2 k \tau }{\gamma
   }\right)-\cosh \left(\frac{2 \tau  u}{\gamma }\right)},\\ 
     y_{c}(t) = \frac{2 \left(\sinh \left(\frac{k t}{\gamma }\right) \cosh \left(\frac{t u}{\gamma }\right) \left(x_{\tau} \sinh \left(\frac{k \tau }{\gamma }\right) \cosh
   \left(\frac{\tau  u}{\gamma }\right)+y_{\tau} \cosh \left(\frac{k \tau }{\gamma }\right) \sinh \left(\frac{\tau  u}{\gamma }\right)\right)\right)}{\cosh \left(\frac{2 k \tau }{\gamma
   }\right)-\cosh \left(\frac{2 \tau  u}{\gamma }\right)}\nonumber\\  +\frac{\left(\-\cosh \left(\frac{k
   t}{\gamma }\right) \sinh \left(\frac{t u}{\gamma }\right) \left(x_{\tau} \cosh \left(\frac{k \tau }{\gamma }\right) \sinh \left(\frac{\tau  u}{\gamma
   }\right)+y_{\tau} \sinh \left(\frac{k \tau }{\gamma }\right) \cosh \left(\frac{\tau  u}{\gamma }\right)\right)\right)}{\cosh \left(\frac{2 k \tau }{\gamma
   }\right)-\cosh \left(\frac{2 \tau  u}{\gamma }\right)}.
\end{eqnarray}

From the extremal path $(x_c(t),,y_c(t))$, one can straightforwardly calculate the Lagrangian for the extremal path, and from it, determine the action for the extremal path. For the general case with $x(0) = x_{0}$ and $y(0) = y_{0}$, the action for the extremal path is given by
\begin{eqnarray}
 \mathcal{S}[x_{c}, y_{c}] = \left(x_{0}^2 + y_{0}^2\right) \left(-\gamma k - \frac{1}{4} \gamma \left((k - u) \cosh\left(\frac{\tau (k - u)}{\gamma}\right) - k \cosh\left(\frac{\tau (3k + u)}{\gamma}\right) \right.\right. \nonumber \\
 \left. + u \cosh\left(\frac{\tau (k + 3u)}{\gamma}\right)\right) \csch\left(\frac{\tau (k - u)}{\gamma}\right) \csch^2\left(\frac{\tau (k + u)}{\gamma}\right)+\frac{1}{2} \gamma \csch\left(\frac{\tau (k - u)}{\gamma}\right)  \nonumber 
\end{eqnarray}
\begin{eqnarray}
 \times\left(-\left((k + u) (x_{0} - y_{0}) (x_{\tau} - y_{\tau})\left(\cosh\left(\frac{2k\tau}{\gamma}\right) - \cosh\left(\frac{2\tau u}{\gamma}\right)\right) \csch^2\left(\frac{\tau (k + u)}{\gamma}\right) \right)\right.\nonumber \\
\ - 2 (k - u) (x_{0} + y_{0}) (x_{\tau} + y_{\tau})+ x_{0} y_{0} \left(2 \gamma u - \frac{1}{2} \gamma \left((k - u) \cosh\left(\frac{\tau (k - u)}{\gamma}\right)\right.\right. \nonumber
\end{eqnarray}
\begin{eqnarray}
\left. + u \cosh\left(\frac{\tau (3k + u)}{\gamma}\right) - k \cosh\left(\frac{\tau (k + 3u)}{\gamma}\right)\right) \csch\left(\frac{\tau (k - u)}{\gamma}\right) \csch^2\left(\frac{\tau (k + u)}{\gamma}\right)  \nonumber \\
+\frac{1}{2} \left(\gamma (k + u) (x_{\tau} - y_{\tau})^2 \left(f_{C}(k+u) + 1\right)  +\gamma (k - u) (x_{\tau} + y_{\tau})^2 \left(f_{C}(k - u) + 1\right)\right).
\end{eqnarray}

and with delta initial condition it reduces to
\begin{eqnarray}
    \mathcal{S}[x_{c},y_{c}] =\frac{1}{2} (\gamma  (k+u) (x_{\tau}-y_{\tau})^2 \left(f_C(k + u) +1\right)\nonumber\\ +\gamma  (k-u) (x_{\tau}+y_{\tau})^2 \left(f_C(k - u)+1\right)).
\end{eqnarray}

With the action on the extreme path, one can calculate the conditional probability, given by
\begin{equation}
    P[x_{\tau},y_{\tau},\tau|x_{0},y_{0},0] = \frac{\exp\left(-\frac{1}{4\gamma T}\mathcal{S}[x_{c},y_{c}]\right)}{\int dx_{\tau} dy_{\tau} \exp\left(-\frac{1}{4\gamma T}\mathcal{S}[x_{c},y_{c}]\right)}.\label{A9}
\end{equation}

\section{Characteristic Function and Distribution}\label{appB}
We are interested in investigating the fluctuations of the harmonic potential and the full potential. One thing we can notice is that since both quantities have a quadratic dependence on the positions, the characteristic function will be of the type
\begin{equation}
Z(\lambda) = \frac{\sqrt{\alpha_3}}{\sqrt{\alpha_1\lambda^2 + \alpha_2\lambda + \alpha_3}}.\label{zlambda}
\end{equation}
With this, we can calculate all moments of the distribution by
\begin{equation}
    \langle X^n \rangle = (i)^n\frac{\partial^n Z(\lambda)}{\partial \lambda^n}\bigg|_{\lambda \rightarrow 0 }.
\end{equation}
These result comes from the integral
\begin{equation}
    Z(\lambda) = \langle e^{-i\lambda X(x,y)} \rangle = \int dx_{\tau}dy_{\tau}dx_{0}dy_{0}\; e^{-i\lambda X(x,y)}P(x_{\tau},y_{\tau},x_{0},y_{0}).\label{zlambdaint}
\end{equation}
since $P(x_{\tau},y_{\tau},x_{0},y_{0})$ is a Gaussian distribution together with the chosen initial conditions which is the zero entropy condition, and $X(x,y)$ is the quadratic functional dependence on the positions. 
What will change will be the coefficients $\alpha_i$'s, depending on the structure of the potential. Nevertheless, we can calculate the Fourier transform of this characteristic function, giving the distribution. We need to notice that we have:
\begin{equation}
P(X) = \int \frac{d\lambda}{2\pi} \sqrt{{\alpha_3}}\frac{e^{i\lambda X}}{\sqrt{\alpha_3 - \frac{\alpha_2^2}{4\alpha_1} + \alpha_1\left(\lambda + \frac{\alpha_2}{2\alpha_1}\right)^2}},
\end{equation}
where, for $\alpha_1 > 0$, we complete the square in the denominator to find this Bessel function definition \cite{chatterjee2010exact,paraguassu2023heat}. Performing the integral, we have:
\begin{equation}
P(X) = \frac{2}{\pi} \sqrt{\frac{\alpha_3}{4\alpha_1}}\exp\left({-i X \frac{\alpha_2}{2\alpha_1}}\right)K_0\left(|X|\sqrt{\frac{4\alpha_3\alpha_1 - \alpha_2^2}{4\alpha_1^2}}\right),\label{distbessel}
\end{equation}
where $X$ is the quantity of interest, which can be either the difference in the harmonic potential $\Delta U$ or the difference in the full potential $\Delta V$. This formula only holds for distributions with $\alpha_{1}>0$. For distributions with $\alpha_{1}<0$, one has to calculate the distribution numerically. We will see that, despite the similarity between the two quantities and having the same distribution, the statistical behavior of the moments is different. 

\section{Skewness and kurtosis for the potentials}\label{appC}

The skewness and kurtosis can be expressed analytically, for all potentials fluctuations. Here we show its long and complicated formulas.
\subsection{Harmonic potential}
For the harmonic potential $\Delta U$, the equation for $\mu_{3,\Delta U}$ is given by
\begin{equation}
    \mu_{3,\Delta U} = k^{3}T^{3}\frac{(\alpha_{\Delta U} + \beta_{\Delta U} + \gamma_{\Delta U})}{\delta_{\Delta U}},\label{skewness_harm_expression}
\end{equation}
where
\begin{eqnarray}
    \alpha_{\Delta U} = 8 k \left(4 k^2+7 u^2\right)+11 (k-u)^3 f_{C}^{3}(k - u)\nonumber\\ + (k-u)^2 \left(5 (k+u) f_{C}(k + u) + 38k - 28u\right)f_{C}^{2}(k - u),\\
     \beta_{\Delta U} = (k+u)f_{C}(k+u)(48 k^2+(k+u)f_{C}(k+u)\left(11 (k+u)f_{C}(k+u) + 38k + 28u\right)\nonumber\\+ 56ku + 28u^2),\\
     \gamma_{\Delta U} = (k-u)f_{C}(k - u)(53 k^2+5 (k+u) \left(2 k \sinh \left(\frac{2 \tau  (k+u)}{\gamma }\right)+k+u\right)\nonumber\\\times \csch^2\left(\frac{\tau
    (k+u)}{\gamma }\right)-46 k u+33 u^2),\\
    \delta_{\Delta U} = (k^{2}-u^{2})^3\left(f_{C}(k - u)+1\right)^3 \left(f_{C}(k + u) + 1\right). 
\end{eqnarray}

The kurtosis is given by
\begin{eqnarray}
    \kappa_{\Delta U} = \frac{12(\alpha_{\Delta U}' + \beta_{\Delta U}'\gamma'_{\Delta U})}{(\delta_{\Delta U}')^{2}},\label{curt_harm_expression}
\end{eqnarray}
with
\begin{eqnarray}
    \alpha_{\Delta U}' = (k - u)^4 (4f_{C}(k - u) + 6f_{C}^2(k - u) + 4f_{C}^3(k - u) + f_{C}^4(k - u)),\\
    \beta_{\Delta U}' = \frac{1}{8} \csch^4\left(\frac{\tau  (k+u)}{\gamma }\right)\\ \nonumber \gamma'_{\Delta U} = (\left(9 k^4+28 k^3 u+54 k^2 u^2+28 k u^3+9 u^4\right) \cosh \left(\frac{4 \tau  (k+u)}{\gamma
   }\right)\\ +8 (k+u)^4 \sinh \left(\frac{4 \tau  (k+u)}{\gamma }\right)-4 (k-u)^4 \cosh \left(\frac{2 \tau  (k+u)}{\gamma }\right)+3 (k-u)^4),\\
   \delta_{\Delta U}' = 2 \left(k^2+u^2\right)+(k+u)^2 f_{C}^2(k+u)+(k-u)^2 f_{C}(k - u) \left(f_{C}(k-u)+2\right)\nonumber\\+2 (k+u)^2 f_{C}(k+u).
\end{eqnarray}

\subsection{Non-confining potential}
For the non-confining potential $\Delta V_{nc}$, the equation for the skewness is given by
\begin{eqnarray}
    \mu_{3,\Delta V_{nc}} = T^3 u^3(\alpha_{\Delta V_{nc}} + \beta_{\Delta V_{nc}} + \gamma_{\Delta V_{nc}} + \delta_{\Delta V_{nc}}),\label{nc_skew_expression}
\end{eqnarray}
with
\begin{eqnarray}
    \alpha_{\Delta V_{nc}} = \left(\frac{11}{(k+u)^3 \left(f_{C}(k + u)+1\right)^3}\right),\\
    \beta_{\Delta V_{nc}} = \left(\frac{5}{(u-k) (k+u)^2 \left(f_{C}(k+u)+1\right)^2 \left(f_{C}(k-u)+1\right)}\right),\\
    \gamma_{\Delta V_{nc}} = \left(\frac{5}{(k-u)^2 (k+u) \left(f_{C}(k+u)+1\right) \left(f_{C}(k - u)+1\right)^2}\right),\\
    \delta_{\Delta V_{nc}} = \left(\frac{11}{(u-k)^3 \left(f_{C}(k-u)+1\right)^3}\right).
\end{eqnarray}

The equation for the excess of the kurtosis, as already mentioned in the text, is equal to the harmonic case, that is, Eq.~\ref{curt_harm_expression}, with the same coefficients.

\subsection{Total potential}
For the total potential, the skewness is not long and complicated, so we only insert the excess of kurtosis here. It is given by 
\begin{equation}
    \kappa_{\Delta V} = \frac{12(2 + \alpha_{\Delta V}'\gamma_{\Delta V}' + \beta_{\Delta V}'\delta_{\Delta V}')}{(2 + \alpha_{\Delta V}' + \beta_{\Delta V}')^{2}},\label{tot_kurtosis_expression}
\end{equation}
where
\begin{eqnarray}
    \alpha_{\Delta V}' = f_{C}(k - u)(2 + f_{C}(k - u)),\;\;\beta_{\Delta V}' = f_{C}(k+u)(2 + f_{C}(k+u)),\\
    \gamma_{\Delta V}' = 2 + \alpha_{\Delta V}',\;\;\delta_{\Delta V}' = 2 + \beta_{\Delta V}'.
\end{eqnarray}

\providecommand{\newblock}{}

\end{document}